\documentclass[iop,apj,twoside]{emulateapj}
\usepackage{amsmath,amssymb,amstext}
\usepackage[breaklinks,colorlinks,citecolor=blue,linkcolor=magenta]{hyperref} 

\usepackage[all]{hypcap}
\usepackage{booktabs}
\usepackage{natbib}

\newcommand{\kms}[0]{km~s$^{-1}$}

\shorttitle{Superbubbles Radiating to 10 K}
\shortauthors{Tanner, Cecil, \& Heitsch}

\begin{document}

\title{Starburst Driven Galactic Superbubbles Radiating to 10 K}

\author{Ryan Tanner, Gerald Cecil, and Fabian Heitsch}
\affil{University of North Carolina at Chapel Hill\\
Chapel Hill, NC 27599-3255\\
rjtanner@physics.unc.edu}

\begin{abstract}
Our three-dimensional hydro-dynamical simulations of starbursts examine the formation of superbubbles over a range of driving luminosities and mass loadings that determine superbubble growth and wind velocity. From this we determine the relationship between the velocity of a galactic wind and the power of the starburst. We find a threshold for the formation of a wind, above which the speed of the wind is not affected by grid resolution or the temperature floor of our radiative cooling. We investigate the effect two different temperature floors in our radiative cooling prescription have on wind kinematics and content. We find that cooling to $10$~K instead of to $10^4$~K increases the mass fraction of cold neutral and hot X-ray gas in the galactic wind while halving that in warm H$\alpha$. Our simulations show the mass of cold gas transported into the lower halo does not depend on the starburst strength. Optically bright filaments form at the edge of merging superbubbles, or where a cold dense cloud has been disrupted by the wind. Filaments formed by merging superbubbles will persist and grow to $>400$ pc in length if anchored to a star forming complex. Filaments embedded in the hot galactic wind contain warm and cold gas that moves $300-1200$ \kms~slower than the surrounding wind, with the coldest gas hardly moving with respect to the galaxy. Warm and cold matter in the galactic wind show asymmetric absorption profiles consistent with observations, with a thin tail up to the wind velocity.
\end{abstract}
\keywords{galaxies: evolution --- galaxies: nuclei --- galaxies: starburst --- ISM: bubbles --- ISM: jets and outflows --- hydrodynamics}

\section{Introduction\label{sec:intro}}
A galactic wind (GW) is a key phase in the gas feedback cycle of galaxies \citep{1990ApJS...74..833H,1994ApJ...427...25S,2001ApJ...560..599A}.
Yet, uncertain coupling of GW to the multi-phase interstellar medium (ISM) obscures how galaxy structure determines the evolution of the wind as its flow alters the ISM.
Models cannot yet fully predict how often and under what circumstances GWs form, and their ultimate impact on galactic evolution.

\citet{1985Natur.317...44C} made the first analytic model of how stellar winds from multiple stars can merge to completely alter the ISM.
Over the first few Myr of a starburst, OB star winds inflate bubbles of hot, low density, metal enriched gas.
Expanding bubbles shock and compress the ISM, then merge as a ``superbubble" of radius $>0.1$ kpc \citep{2013PASA...30...25D} that is powered first by OB and WR-star winds then SNe~II. 
The superbubble can expand to exceed the scale height of the galaxy, potentially ``blowing out'' its metal-enriched gas into the low density halo \citep[the ``champagne effect'',][]{TenorioTagle79} forming a galactic wind. 
See \citet{2005ARA&A..43..769V} and references therein for an extensive overview of GWs.

Various models have been used to investigate the effect of different parameters on starburst driven GWs. \citet{1988ApJ...324..776M} showed that blowout likelihood is proportional to the mechanical luminosity of the starburst, and inversely proportional to the ISM pressure and disk scale height. \citet{1994ApJ...430..511S} concluded that GW development depended on the nature of mass and energy injection in the starburst region. \citet{1996ApJ...468..722S} found that lower average densities in a non-uniform ISM increased bubble size, and that an increase in mass loading decreases the interior temperature of the superbubble. Further work by \citet{1999MNRAS.309..332T} found that a superbubble blowout into the inter-galactic medium (IGM) depends heavily on the power of the nuclear starburst. \citet{StricklandStevens} studied how ISM distribution, starburst characteristics and mass loading affect X-ray emission, and mass and energy transport into the IGM by the GW. \citet{2009ApJ...698..693F} and \citet{2009ApJ...697.2030S} simulated starbursts with different mass loadings and mechanical luminosities and determined the relationship to mass flow rates and GW terminal velocities.
\citet{CooperI} found that a blowout is channeled by the scale height, density, and pressure of the ambient disk ISM. 
\citet{2013MNRAS.430.3235M} investigated the dependence of GW evolution on the environment at the base of the GW and determined that optical filament formation depends on the clumpiness of the starburst region. \citet{Creasey} showed that higher gas surface density and lower gas fraction should make faster GWs.

Simulations of starburst driven GWs have included radiative cooling, but few have examined the effects of cooling below $10^4$ K. Early work by \citet{1988ApJ...324..776M,1989ApJ...337..141M,1994ApJ...430..511S} and \citet{1996ApJ...468..722S} approximated cooling with a power-law relation down to $10^{5}$ K. 
Subsequent studies have used the cooling tables of \citep{1993ApJS...88..253S,1976ApJ...204..290R,1986RvMP...58....1S} down to $10^4$ K. 
\citet{StricklandStevens}, and \citet{SutherlandBicknell} addressed X-ray emission but not emission from cold gas and thus did not include cooling below $10^4$ K. \citet{2009ApJ...697.2030S} used post processing to calculate emission but did not include cooling in their simulations. 
\citet{CooperI} considered H$\alpha$ emission and X-rays, but were matching optical data.
\citet{Creasey} argued that energy loss below 8,000 K is insignificant and does not affect GW formation.
\citet{2006ApJ...653.1266J} used a parameterized cooling curve  \citep{1972ARA&A..10..375D} below $10^4$ K to examine formation of cold dense clouds near supernovae. 
\citet{2009ApJ...698..693F} found that cooling below $10^4$ K does not affect gas outflow kinematics. Because the effect of low temperature cooling has not been detailed we therefore compare the effects of cooling below $10^4$ K versus cooling terminated at $10^4$ K on wind dynamics, emission and content.

Our simulations test these expectations over the first 1.5 Myr following a single instantaneous starburst. This is sufficient time for the superbubble to blow out of the disk and form a GW. A study of how the GW interacts with the galactic halo would require a longer simulation time and a more extensive box size than we consider here.
For consistency with previous studies of starbursts \citep{CooperI,2009ApJ...697.2030S,2013MNRAS.430.3235M}, we fix galaxy size and shape at M82 values to focus on mechanical luminosity and mass loading as a superbubble forms.
We will show clear GW thresholds for both parameters.

GWs are traced by filamentary optical \citep{1988Natur.334...43B,1994ApJ...433...48V,1998ApJ...493..129S,1999ApJ...510..197D} and X-ray emission \citep{1997A&A...320..378S,2002ApJ...568..689S}; and molecular \citep{2002ApJ...580L..21W} and atomic \citep{2002ApJ...570..588R,2005ApJS..160..115R} absorption.
Structures in the emitting bands are tightly correlated, e.g.\
\citet{2002ApJ...576..745C} combined \textit{Chandra}, \textit{HST}, and \textit{VLA} datasets to characterize the environment and emitting filament towers of the GW in NGC 3079. 
Those authors conclude that the towers form at the edge of the starburst and are remnants of the ISM propelled by the starburst, not from condensed wind. 
To determine how filaments can be used as tracers of wind dynamics we therefore consider filaments over temperatures that span from X-ray to molecular emission. Most previous work \citep{2009ApJ...698..693F,2009ApJ...697.2030S,2013MNRAS.434.3572R,2014MNRAS.441..431S} simulated starbursts in 2D with more recent work focusing on 3D \citep{CooperI,2013MNRAS.430.3235M,Creasey}. We perform all our simulations in 3D to fully explore the formation and structure of filaments in the GW.

With a simplified ISM substrate plus a fractal distribution of denser clouds, our simulations (\S\ref{sec:problem}) explore ranges of two parameters of a nuclear starburst --- its mechanical luminosity and its mass loading of a GW --- that form and evolve a GW.
We compare two sets of simulations with different temperature cutoffs for radiative cooling ($10^4$ K vs. $10$ K) to examine the effects on GW properties (\S\ref{sec:blowout}), and the outcomes of cooling to $10$~K on the multi-band emission (\S\ref{sec:blowout:Emis}). 
We are motivated in part by Herschel Observatory \citep[e.g.][]{2015ApJ...804...46M},  and ground-based studies \citep[e.g.][]{2002ApJ...580L..21W} that map cold filaments $>1$ kpc above nearby starbursts. We therefore consider the kinematics (\S\ref{sec:fila}) and absorption profiles (\S\ref{sec:lines}) of filaments.

\section{Numerical Methods}\label{sec:problem}
We integrate numerically the inviscid hydrodynamical equations with the public Athena code \citep{Stone-Athena}. 
The Appendix  lists our modifications to improve code stability as large pressure and density variations are encountered during cooling to low temperatures. 

\subsection{Gravitational Potential and Initial Velocity Field}\label{sec:problem:grav}
Following \citet{CooperI} and \citet{StricklandStevens} we model the stellar gravitational potential as a combined disk and bulge. The disk, with mass $M_{\rm disk}$, radial scale size $a$, and vertical scale size $b$ is modeled as a Plummer-Kuzmin potential \citep{1975PASJ...27..533M}
\begin{equation}\label{phidisk}
\Phi_{\rm disk}(r,z)=-\frac{GM_{\rm disk}}{\sqrt{r^{2}+(a+\sqrt{z^{2}+b^{2}})^{2}}}
\end{equation}
The spheroidal bulge $\Phi_{\rm ss}(R)$ is modeled as a King model, 
\begin{equation}\label{phiss}
\Phi_{\rm ss}(R)=-\frac{GM_{\rm ss}}{r_{0}}\left[\frac{\ln\left[(R/r_{0})+\sqrt{1+(R/r_{0})^{2}}\right]}{(R/r_{0})}\right],
\end{equation}
with $R=\sqrt{r^{2}+z^{2}}$, radial scale size $r_{0}$, and mass $M_{\rm ss}$. The total potential is $\Phi_{\rm tot} = \Phi_{\rm disk} + \Phi_{\rm ss}$ using Equations \ref{phidisk} and \ref{phiss}. We neglect the contribution of the dark matter halo since our simulation only covers the central 1 kpc. In that region matter is baryon dominated \citep{2011MNRAS.414.2446M}. The disk gas is initially rotating at azimuthal velocity 
\begin{equation}
v_{\phi}(r,z)=e_{\rm disk}\exp(-|z|/z_{\rm rot})\,\left(r\frac{\partial\Phi_{tot}(r,0)}{\partial r}\right)^{1/2}
\end{equation}
Here $e_{\rm disk}$ is the ratio azimuthal to Keplerian velocity. Table \ref{tab:param} lists simulation parameter values. The parameters have been chosen to match the rotation curve of M82 \citep{StricklandStevens,CooperI}. All boundaries in the simulation box are outflow boundaries. Any gas that reaches a boundary due to the initial rotation is lost.

\subsection{Gas Thermal Balance}\label{sec:problem:cool}
The Athena code implements thermal physics as an external source term in the total energy equation. 
To range over the $10<T<10^{8}$~K anticipated in our simulations, we combined tabulated cooling curves for solar metallicity \citep{1993ApJS...88..253S} with the low-temperature photoelectric heating (eq. \ref{eq:heating}) and cooling (eq. \ref{eq:coolingfunction}) of \citet{Koyama} based on \citet{1995ApJ...443..152W}, with appropriate corrections by \citet{KoyamaII}. \citet{2015ApJ...802...99K} have used a similar implementation of heating and cooling in Athena. 
The rate of energy change \citep{Field} is
\begin{equation}
\mathcal{L}=n(\Gamma-n\Lambda(T)).\label{eq:heatingandcooling}
\end{equation}
with heating
\begin{equation}\label{eq:heating}
\Gamma =\left\{ \begin{array}{lr}
2\times10^{-26} \text{~erg~cm}^{-3}~\text{s}^{-1}& : T < 10^{4} \text{ K}\\
0 & : T > 10^{4} \text{ K}
\end{array}\right.
\end{equation}
and cooling where $T< 10^{4}$~K 
\begin{equation}
\frac{\Lambda(T)}{\Gamma}=10^{7}\exp\left(\frac{-118400}{T+1000}\right)+0.014\sqrt{T}\exp\left(\frac{-92}{T}\right)\text{cm}^{3}.\label{eq:coolingfunction}
\end{equation}
For $10^{4}<$ T $<10^{8.5}$ K, we use piecewise power-law fits to the tabulated cooling for collisional ionization equilibrium at solar metallicity from \citet{1993ApJS...88..253S}. Although we do not anticipate temperatures above $10^{8}$~K, for completeness we include emission through bremsstrahlung above $T>10^{8.5}$ K using \citep{1986rpa..book.....R}
\begin{equation}
\Lambda=2.1\times10^{-27}T^{1/2}n^{2}Z^{2}.
\end{equation}
We use Eq.\ \ref{eq:heatingandcooling} to calculate cell emissivity and sum radiative losses along a chosen column to calculate gas emission. 
We separate emission into bands for cold gas, H$\alpha$, soft X-ray, mid X-ray and hard X-ray emission. Table \ref{tab:body:TempEmis} gives temperature ranges for the bands.

We run two sets of simulations with different cutoff temperatures where cooling is applied, one with cooling only applied when gas temperature $> 10^4$ K, the other with cooling applied down to $10$ K. In both cases we impose a  temperature floor at $10$ K.
\begin{table}
\begin{center}
\caption{Definition of gas temperature ranges\label{tab:body:TempEmis}}
\begin{tabular}{ll} \toprule
Band & Range \\
\midrule
  Cold Gas & $<$ 100 K\\
  H$\alpha$ & 5e3-4e4 K\\
  Soft X-Ray & 0.5-3.0 keV \\
  Mid X-Ray & 3.0-10.0 keV\\
  Hard X-Ray & $>$ 10.0 keV \\
\bottomrule
\end{tabular}
\end{center}
\end{table}

\subsection{Initial Conditions of the ISM}\label{sec:problem:ini}
To generate a realistic initial ISM, we multiply a smooth background against a fractal density distribution to mimic embedded clouds.

\subsubsection{Smooth ISM}\label{sec:problem:ISM}
Densities in the computational domain are a combination of halo and disk distributions given by
\begin{multline}\label{eq:smoothdens}
n_{\rm halo}(r,z) =  n_{\rm halo}(0,0)\times \\
\exp\left[-\frac{\Phi_{\rm tot}(r,z)-e_{\rm halo}^{2}\Phi_{\rm tot}(r,0)-(1-e_{\rm halo}^{2})\Phi_{\rm tot} (0)}{c_{\rm s,halo}^{2}}\right], \\
n_{\rm disk}(r,z) =  n_{\rm disk}(0,0)\times \\
\exp\left[-\frac{\Phi_{\rm tot}(r,z)-e_{\rm disk}^{2}\Phi_{\rm tot}(r,0)-(1-e_{\rm disk}^{2})\Phi_{\rm tot}(0)}{\sigma_{t}^{2}+c_{\rm s,disk}^{2}}\right],
\end{multline}
central density $n(0,0)$, sound speed $c_{s,disk} = \sqrt{k_BT_{disk}/m_H}$ that sets the scale height of each density profile, and $e_{\rm disk,halo}$ the ratio of azimuthal to Keplerian velocity.
The turbulence parameter $\sigma_{t}$ helps to form a thick disk without raising its temperatures artificially \citep[see][]{CooperI}.

\begin{figure}[!ht]
\centering
\includegraphics[width=0.5\textwidth]{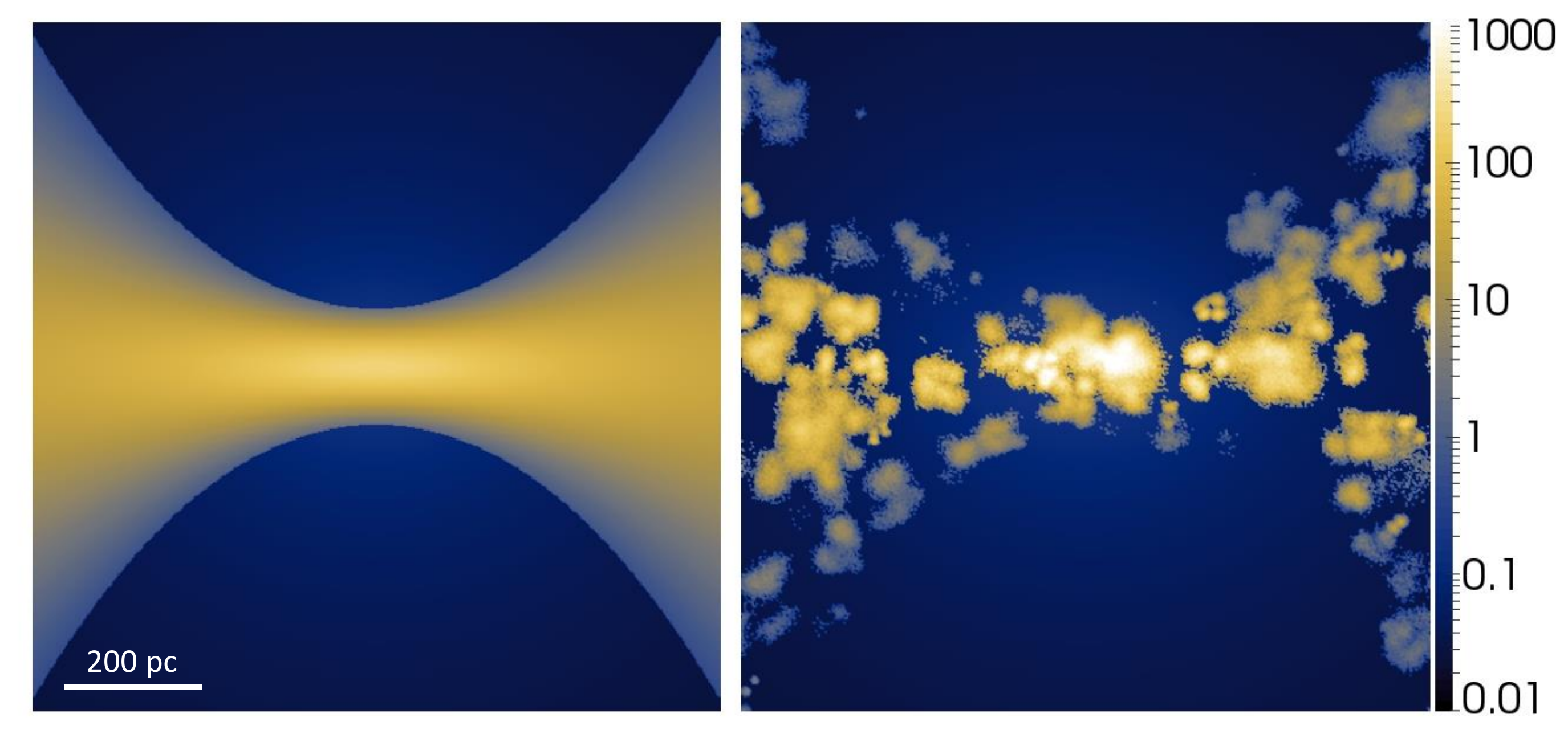}
\protect\caption{XZ plane slice of gas density ($n(r,z)$ in cm$^{-3}$) scaled logarithmically. {\em Left:}
Smooth disk before adding fractal clouds. 
{\em Right:} The disk with fractal clouds.}
\label{fig:test:inidisk} 
\end{figure}
\begin{table}
\begin{center}
\caption{Parameters used for simulation setup.\label{tab:param}}
\footnotesize
\begin{tabular}{lll} \toprule
Symbol & Value & Property \\
\midrule
\multicolumn{3}{l}{Parameters used for initial gas distribution.} \\ \midrule
  $n_{\rm halo}(0,0)$ & $0.2$ particles/cm${}^3$ & Central halo density \\
  $n_{\rm disk}(0,0)$ & $100$ particles/cm${}^3$ & Average density in starburst \\
  $T_{\rm halo}$ & $5.0\times 10^6$ K & Halo temperature \\
  $T_{\rm disk}$ & $1.0\times 10^4$ K & Average disk temperature \\
  $\sigma_t$ & $60$ \kms & Turbulence parameter for disk \\
  $e_{\rm disk}$ & $0.95$ & Rotation ratio (disk) \\
  $e_{\rm halo}$ & $0.00$ & Rotation ratio (halo) \\ \midrule
\multicolumn{3}{l}{Parameters used for the starburst.} \\ \midrule
  $R_{\rm sb}$ & $150$ pc & Starburst radius\\
  $H_{\rm sb}$ & $60$ pc & Starburst height\\ \midrule
\multicolumn{3}{l}{Parameters used for the gravitational potential.} \\ \midrule
  $M_{\rm ss}$ & $6\times 10^8 M_\sun$ & Stellar spheroid mass \\
  $M_{\rm disk}$ & $6\times 10^9 M_\sun$ & Stellar disk mass \\
  $r_0$ & $350$ pc & Stellar spheroid radial scale size \\
  $a$ & $150$ pc & Disk radial scale size \\
  $b$ & $75$ pc & Disk scale size \\
  $z_{\rm rot}$ & $500$ pc & Rotational scale height \\ 
\bottomrule
\end{tabular}
\end{center}
\end{table}

\subsubsection{Fractal Clouds}\label{sec:problem:frac}
A ``cloudy'' ISM is mimicked by a fractal density distribution, multiplied against the smooth background disk density
\begin{equation}
n(r,z)=n_{\rm halo}(r,z)+n_{\rm disk}(r,z)N(r,z)
\end{equation}
with $N(r,z)$ the fractal density fraction of each grid cell. 
To make a fractal density distribution we generate a set of individual fractal clouds following \citet[\S2]{Mathis} with modifications. 
We repeat the Mathis et al.\ approach for a single fractal cloud $n_{c}$ times (see below), but with the constraint that first-level points must fall a distance of $\ge L/4$ from the edge of the box. We place each cloud within the computational domain and repeat for $n_{c}$ fractal clouds with a scale length chosen at random between $50<L<150$~pc. Each cloud is placed semi-randomly on the computational grid to avoid excessive overlap. 
To set $n_{c}$, we repeat until the average fractal density of the grid equals the density of a single cloud.

For models with cooling applied only when $T>10^{4}$~K, we set the disk pressure using $P_{\rm disk}(r,z) = n_{\rm disk}(r,z)c^2_{\rm s,disk}$. 
For models with cooling applied down to $T>10$~K, the heating/cooling function sets the disk to thermal equilibrium (see \S\ref{sec:problem:cool}). In this case the disk pressure is $P_{\rm disk}(r,z) = n_{\rm disk}(r,z)k_{B}T_{\rm TE}$. 
In both cases when $T>3\times10^{4}$~K, cells are set to halo densities and pressures only. 
This prescription is given as,
\begin{equation}\label{eq:pressure}
P(r,z)=\left\{ \begin{array}{lcr}
n_{\rm halo}(r,z)c_{\rm s,halo}^{2}+P_{\rm disk}(r,z) &:&<3\times10^{4}\text{ K}\\
n_{\rm halo}(r,z)c_{\rm s,halo}^{2} &:&>3\times10^{4}\text{ K}
\end{array}\right.
\end{equation}
We use the adiabatic exponent $5/3$ and mean molecular weight $1$. 

A file containing all fractal points was generated with $512^{3}$ grid cells. 
It initialized all models, being coarsened for lower resolution models so that the same initial density distribution was used for all models.

\subsection{Starburst}\label{sec:problem:star}
We model a spheroidal central starburst using
\begin{equation}
1>\frac{(x^{2}+y^{2})}{(R_{\rm sb}^{2})}+\frac{(z^{2})}{(H_{\rm sb}^{2})},
\end{equation}
of radius $R_{\rm sb}$ and height $H_{\rm sb}$. 
At each time step we inject mass and energy into the starburst volume at rates $\dot{M}$ and $\dot{E}$. Each cell in the starburst region is injected with mass and energy proportional to that cell's fraction of the total initial ISM mass within the starburst volume. At each timestep we calculate the change in the mass ($dM$) and energy ($dE$) of each cell inside the starburst using
\begin{align}
\frac{dM}{dtdV_{cell}} & =\frac{\dot{M}n_{ini}}{\int n_{ini}dV_{SB}}\\
\frac{dE}{dtdV_{cell}} & =\frac{\dot{E}n_{ini}}{\int n_{ini}dV_{SB}}.
\end{align}
Here $dV_{cell}$ is the cell volume, $n_{ini}$ is the initial density of the cell. The energy injection rate ($\dot{E}$) for the entire starburst is directly related to the mechanical luminosity of the stars by
\begin{equation}\label{eq:mechlum}
\dot{E} = \epsilon L_{\star},
\end{equation}
where $\epsilon$ is the thermalization efficiency and $L_{\star}$ the mechanical luminosity \citep{2005ARA&A..43..769V}. The exact value of $\epsilon$ depends on the local environment of the stars in the starburst and is time dependent \citep{2003ApJ...594..888F,2005ARA&A..43..769V,2009ApJ...697.2030S,2015ApJ...802...99K}. \citet{2003ApJ...594..888F} show that the thermalization efficiency varies over time ranging from 0.1 immediately after star formation to $\sim 0.01$. \citet{2009ApJ...697.2030S} mention that 0.1 is the practical lower limit for the thermalization efficiency and conclude that a proper value for M82 ranges from 0.3 to just shy of 1.0. While \citet{2015ApJ...802...99K} find a thermalization efficiency ranging from 0.1 to 1.0, but it is highly time dependent, with rapid shifts between values of 1.0 and 0.1-0.3. For simplicity we set $\epsilon=1$.

Using Starburst99 population synthesis models \citep{Leitherer1999} we relate a range of energy injection rates to the total mass of a single instantaneous starburst (SIB). Parameter values given in \S\ref{sec:blowout:models} yield a mass scale of $5\times10^{6}<M<1\times10^{8}M_{\sun}$. \citet{2008A&A...484..711B} give a total mass for the starburst in M82 of $\sim4\times10^{7}M_{\sun}$. Thus our simulations exceed the range of SIBs comparable in mass to the starburst in M82 to adequately investigate the limit of a superbubble blowout.

Supernovae do not contribute substantially until 3-6 Myr after the burst begins. We therefore only consider the contribution from stellar winds and molecular clouds. Like most high-resolution simulations \citep{1996ApJ...463..528S,CooperI,2009ApJ...697.2030S}, we combine contributions of stellar mass loss with that ablated from cold molecular clouds that are unresolved in our simulations as given in Equation \ref{eq:massload}.
\begin{equation}\label{eq:massload}
\dot{M} = \dot{M}_{\star}+\dot{M}_{cold} = \beta\dot{M}_{\star},
\end{equation}
with $\beta$ the mass loading factor. ($\dot{M}$ is called the central mass loading by \citet{1996ApJ...463..528S}, or the mass injection rate by \citet{CooperI} and \citet{2009ApJ...697.2030S}. We call it the mass loading rate.)

\citet{2009ApJ...698..693F} explored mass loading rates ranging from 1.7 M$_\sun$~yr$^{-1}$ to 120 M$_\sun$~yr$^{-1}$. \citet{2009ApJ...697.2030S} explored a much smaller range and determined a mass flow rate corresponding to M82 to be $1.4 \lesssim \dot{M} \lesssim 3.6$ M$_\sun$~yr$^{-1}$. We choose mass loading values, given in \S\ref{sec:blowout:models}, that are similar to \citet{2009ApJ...697.2030S}. This corresponds to values $2 \lesssim \beta \lesssim 15$ for the most energetic starbursts and $35 \lesssim \beta \lesssim 242$ for the smallest.

\begin{figure}[!ht]
\centering 
\includegraphics[width=0.53\textwidth]{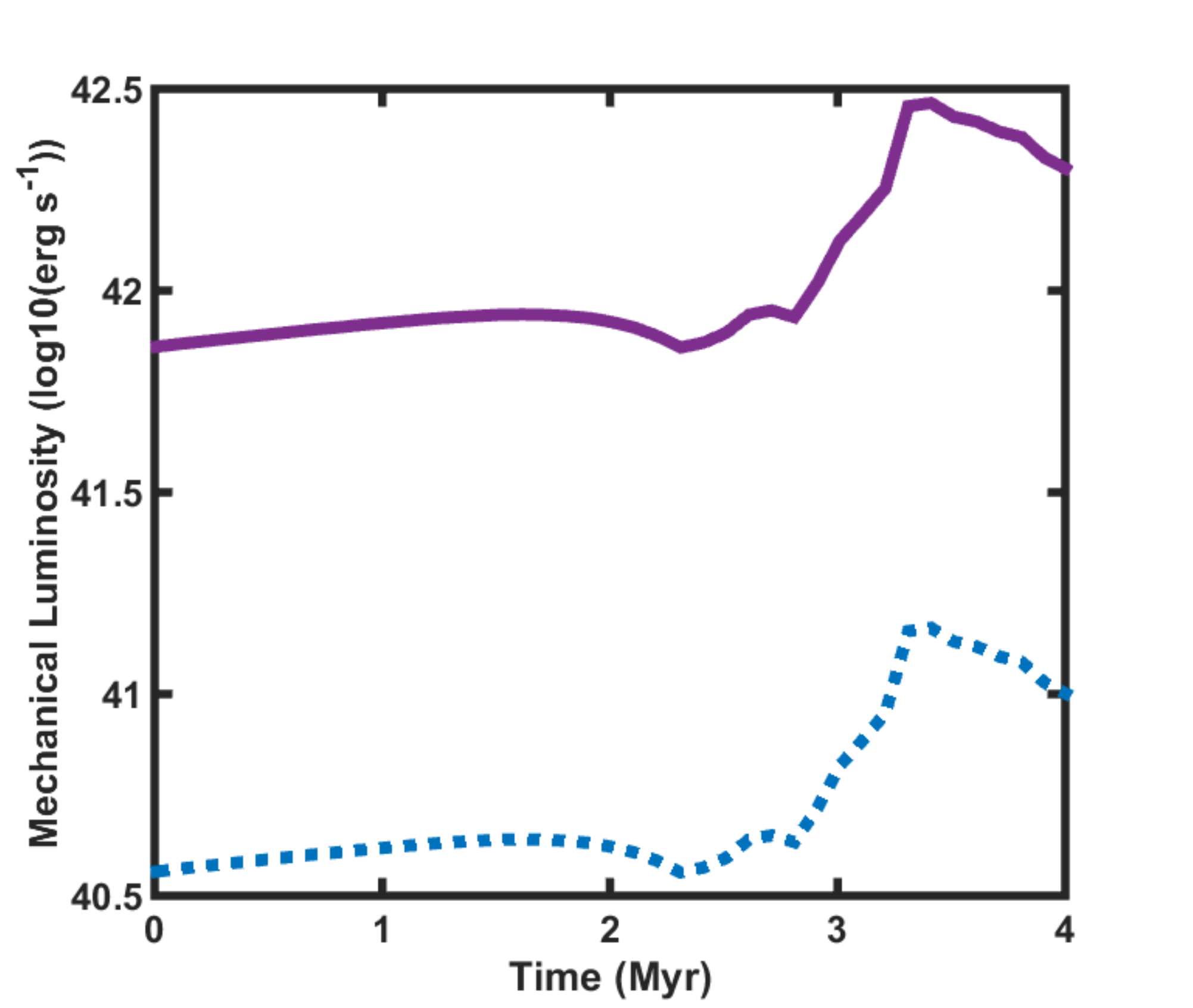}
\protect\caption{$L_{\star}$ (erg~s$^{-1}$) for starbursts of $5\times10^6\text{M}_{\sun}$ (dashed) and $1\times10^8\text{M}_{\sun}$ (solid) to match the range described in \S\ref{sec:blowout:models}, using Starburst99 population synthesis models \citep{Leitherer1999}.
All our analysis is done at 1.5 Myr when all models have achieved a steady-state solution, but before supernovas explode. Therefore we only consider a constant energy input.}
\label{fig:test:ML} 
\end{figure}

Because $L_{\star}$ and $\dot{M}_{\star}$ calculated by Starburst99 are roughly constant for the first 3 Myr of a burst (Fig.\ \ref{fig:test:ML}), we inject mass and energy into the ISM at constant rates. In our models, energy is injected only as as internal energy, not kinetic energy.

\subsection{Model Parameters}\label{sec:blowout:models}
We simulate within a cube 1 kpc on a side divided into $128^3$, $256^3$, or $512^3$ fixed cells with spatial resolution $7.8$, $3.9$, or $2.0$~pc respectively. 
For our low resolution models we vary $0.5 \leq \dot M \leq 3.5$~M$_\sun$~yr$^{-1}$ in steps of $0.5$ M$_\sun$ yr$^{-1}$, and $5\times10^{40} \leq \dot E \leq 1\times10^{42}$~erg~s$^{-1}$ in steps of $0.25$ dex. We have nine medium resolution models ranging from $1.0 \leq \dot M \leq 2.0$~M$_\sun$~yr$^{-1}$ and $1\times10^{41} \leq \dot E \leq 5\times10^{41}$~erg~s$^{-1}$ with another medium resolution model at $\dot{M} = 1.0$~M$_\sun$~yr$^{-1}$, $\dot{E} = 1\times10^{42}$~erg~s$^{-1}$. These ranges were chosen to straddle the transition from blowout to no blowout. We have two high resolution models which use $\dot{M} = 1.5$~M$_\sun$~yr$^{-1}$, $\dot{E} = 2.5\times10^{41}$~erg~s$^{-1}$ and $\dot{M} = 1.0$~M$_\sun$~yr$^{-1}$, $\dot{E} = 1\times10^{42}$~erg~s$^{-1}$. The former was chosen to study a low energy GW, while the latter was chosen to study a high energy GW and for comparison to \citet{CooperI} who use the same mass and energy injection rates.
Model numbers denote grid resolution, $\dot M$, $\dot E$ and cooling used. 
Models starting with ``M1'', ``M2'' or ``M5'' correspond to $128^3$, $256^3$, and $512^3$ cells respectively. 
Postfix indicies designate $\dot M$ and $\dot E$ respectively, see Table \ref{tab:test:tests} column 1.
T4 models cool to $10^4$~K, T1 models to 10 K.
To summarize our nomenclature, model ``M1\_34T4'' has $128^3$ cells with $\dot M = 1.5$~M$_\sun$~yr$^{-1}$, $\dot E = 2.5\times10^{41}$~erg~s$^{-1}$, and cooling limited to $T>10^{4}$~K.

\begin{table}
\begin{center}
\caption{\label{tab:test:tests}$\dot M$ and $\dot E$ used for Fig.\ 3. Index refers to model number. First index in model number corresponds to $\dot M$, second to $\dot E$.}
\begin{tabular}{ccc} \toprule
Index & $\dot{M} (M_\sun yr^{-1})$ & $\dot{E} (erg~s^{-1})$ \\
\midrule
  1 & $0.5$ & $5.0e40$ \\
  2 & $1.0$ & $7.5e40$ \\
  3 & $1.5$ & $1.0e41$ \\
  4 & $2.0$ & $2.5e41$ \\
  5 & $2.5$ & $5.0e41$ \\
  6 & $3.0$ & $7.5e41$ \\
  7 & $3.5$ & $1.0e42$ \\
\bottomrule
\end{tabular}
\end{center}
\end{table}

We ran the 49 combinations of $\dot M$ and $\dot E$ in Table \ref{tab:test:tests} with $128^3$ cells, ten combinations with $256^3$, and two with $512^3$. Each model was run twice, once with cooling to $10^4$~K and then with cooling to 10 K, for a total of 122 models.

\section{Blowout Conditions and Structure}\label{sec:blowout}
\subsection{Wind Structure}\label{sec:blowout:Wind}
\begin{figure*}[h]
\centering
\includegraphics[width=1.03\textwidth]{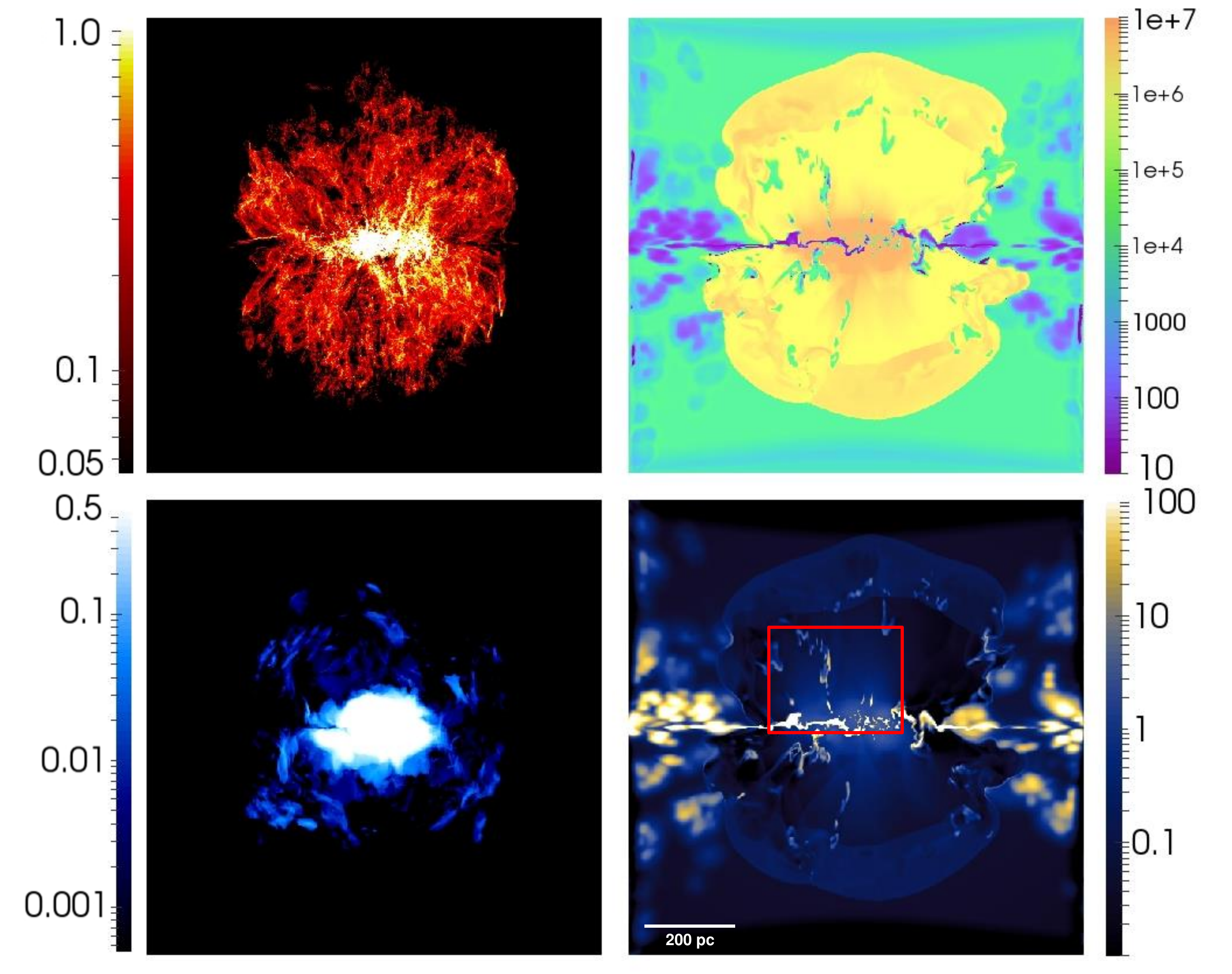}
\protect\caption{A slice in the yz plane through the center of the galaxy for model M5\_34T1 at $1.5$~Myr. 
Clockwise from top left: H$\alpha$ emission ($\log$ erg~s$^{-1}$~cm$^{-2}$) and temperature ($\log$~K), density ($\log$~cm$^{-3}$), and soft X-ray emission scaled as $\log$(~erg~s$^{-1}$~cm$^{-2}$). Red box in bottom right image indicates the zoomed-in region of Figure~\ref{fig:test:DV34}.}
\label{fig:test:dTHaXR34}
\end{figure*}

\begin{figure*}[h]
  \centering 
  \includegraphics[width=1.03\textwidth]{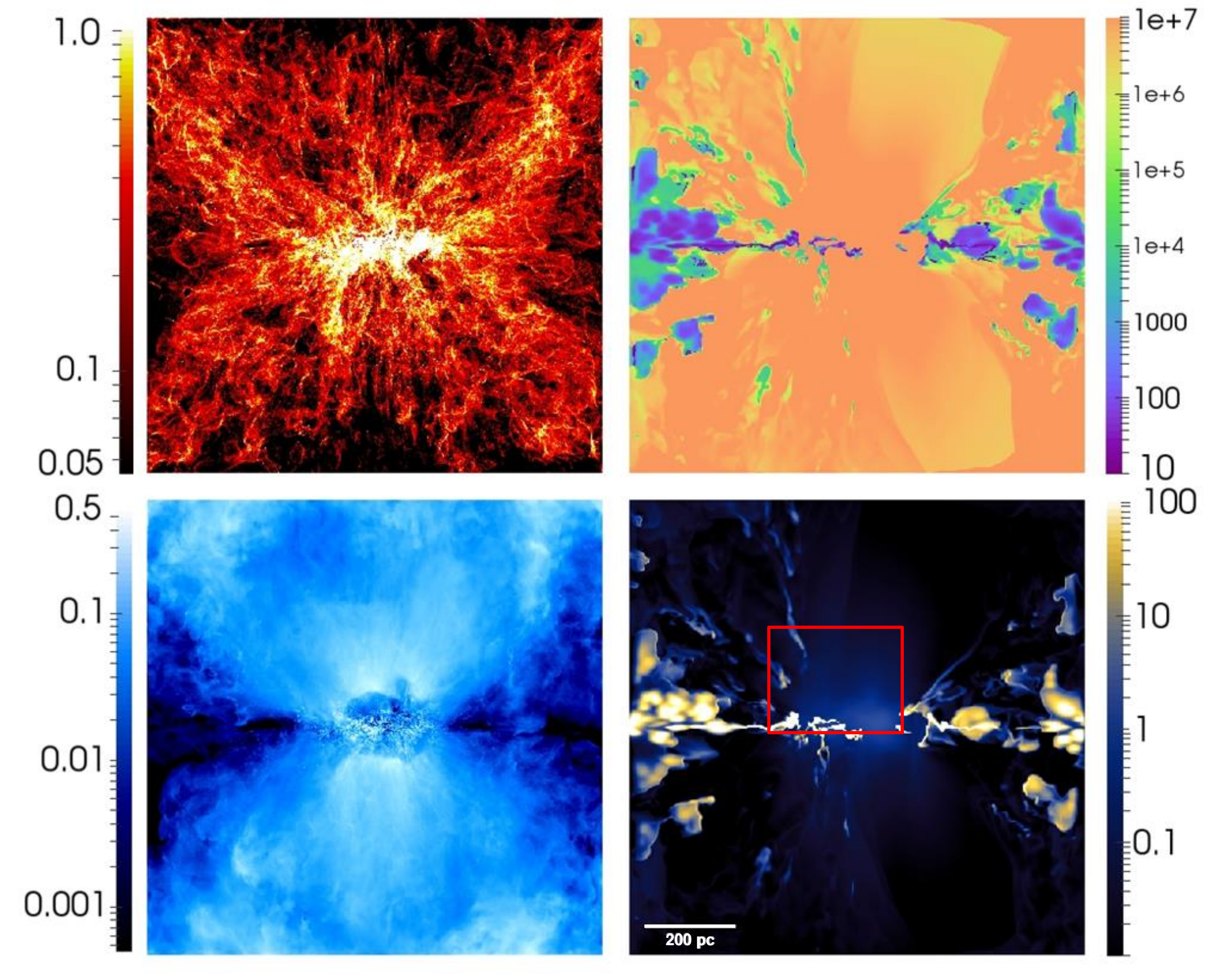}
  \protect\caption{Same as Figure 4, but now for model M5\_27T1 at $1.5$~Myr. Red box in bottom right image indicates the zoomed-in region of Figure~\ref{fig:test:DV27}.}
\label{fig:test:dTHaXR27} 
\end{figure*}

Figures \ref{fig:test:dTHaXR34} and \ref{fig:test:dTHaXR27} show a ``typical'' GW in our highest resolution models (M5\_34T1 and M5\_27T1).
They plot at 1.5 Myr a yz-slice of temperature and density together with column integrated H$\alpha$ and soft X-ray emission.  
The mass and energy injection rates of model M5\_27T1 powers a GW of terminal velocity $\sim1420$~\kms. 
Our M5\_34T1 model with a quarter the energy injection but 50\% higher mass injection rate still forms a GW but with a terminal velocity of $\sim540$~\kms. 
After $1.5$~Myr, model M5\_34T1 has accumulated enough energy to blow out (Fig.~\ref{fig:test:dTHaXR34}) but insufficient to clear the entire volume as model M5\_27T1 does.

Models that blow out have a hot ($\gtrsim10^{6}$~K) free-wind region where the velocity is set by $\dot E$ and $\dot M$. 
Embedded in the free wind are dense ($>10$ particle cm$^{-3}$) filaments of warm and cold gas ($<5000$~K) surrounding dense cores ($>100$ particle cm$^{-3}$)  that have been swept up by the wind. 
These filaments are discussed in \S\ref{sec:fila}. 
The swept-up gas substrate is shock heated to $\gtrsim10^{7}$~K and surrounds the free wind as a shell. 

\subsection{Outflow Wind Speed}\label{sec:blowout:WindSpeed}
\begin{figure}[!ht]
\centering
\includegraphics[width=0.5\textwidth]{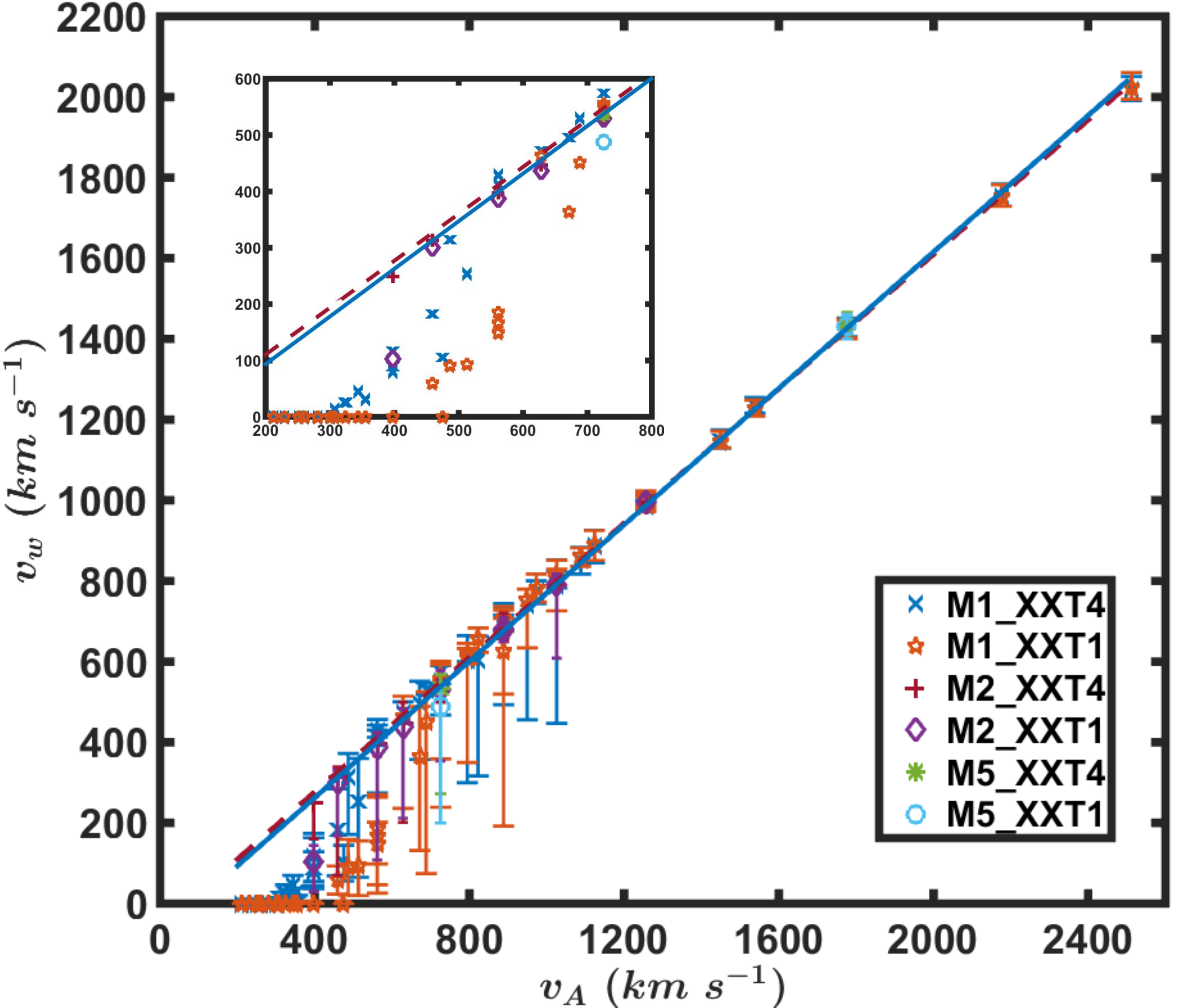}
\protect\caption{Analytical wind speed (from Eq.\ \ref{eq:body:MErel2}) vs.\ simulated vertical wind velocity ($v_w$) $100$~pc above the disk plane at 1.5 Myr. 
Linear fits are shown for all simulations having $v_{A}>500$ \kms. 
Inset: Close up of the break where analytical wind speed deviates from the simulated value.}
\label{fig:test:Windwind}
\end{figure}

The analytic terminal wind speed of a blowout is related to $\dot E$ and $\dot M$  \citep[see][based on \citet{1977ApJ...218..377W} and \citet{1987ApJ...317..190M}]{2009ApJ...698..693F} as 
\begin{equation}
v_A\equiv\left(2\frac{\int\dot{E}dt}{\int\dot{M}dt}\right)^{1/2}.\label{eq:body:MErel2}
\end{equation}
It is related to the simulated wind speed $(v_w)$ by
\begin{equation}
v_{w}=\xi^{1/2} v_A\label{eq:body:MErel}
\end{equation}
\citet{2009ApJ...698..693F} give $\xi=5/11\approx0.45$ which is the fraction of $\dot E$ that drives the kinetic energy within a bubble that is embedded in a uniform ISM \citep{1977ApJ...218..377W}. 
For comparison to analytical results, we determine $\xi$ from our model set (Fig.~\ref{fig:test:Windwind}). 
Eq.~\ref{eq:body:MErel} is generally reproduced by our models: T4 models when $\xi=0.650\pm0.007$; T1 models when $\xi=0.68\pm0.03$ for $v_A>600$~\kms.

Escape velocity from the model galaxy is $v_e\approx 490$ \kms.
For $v_A<v_e$, our simulations do not blow out. 
For $v_A > 1.5v_e$, our T4 and T1 series are identical, and increased resolution does not alter the wind speed.
In the transition $v_e < v_A < 1.5v_e$, our T4 models have higher simulated wind speeds than T1 models (Fig.\ \ref{fig:test:Windwind} inset); both deviate from the relation in Eq.\ \ref{eq:body:MErel}. 

\subsection{Emission as Blowout Tracer}\label{sec:blowout:tracer}
\begin{figure*}
\label{fig:test:128all}
\centering
\includegraphics[width=\textwidth]{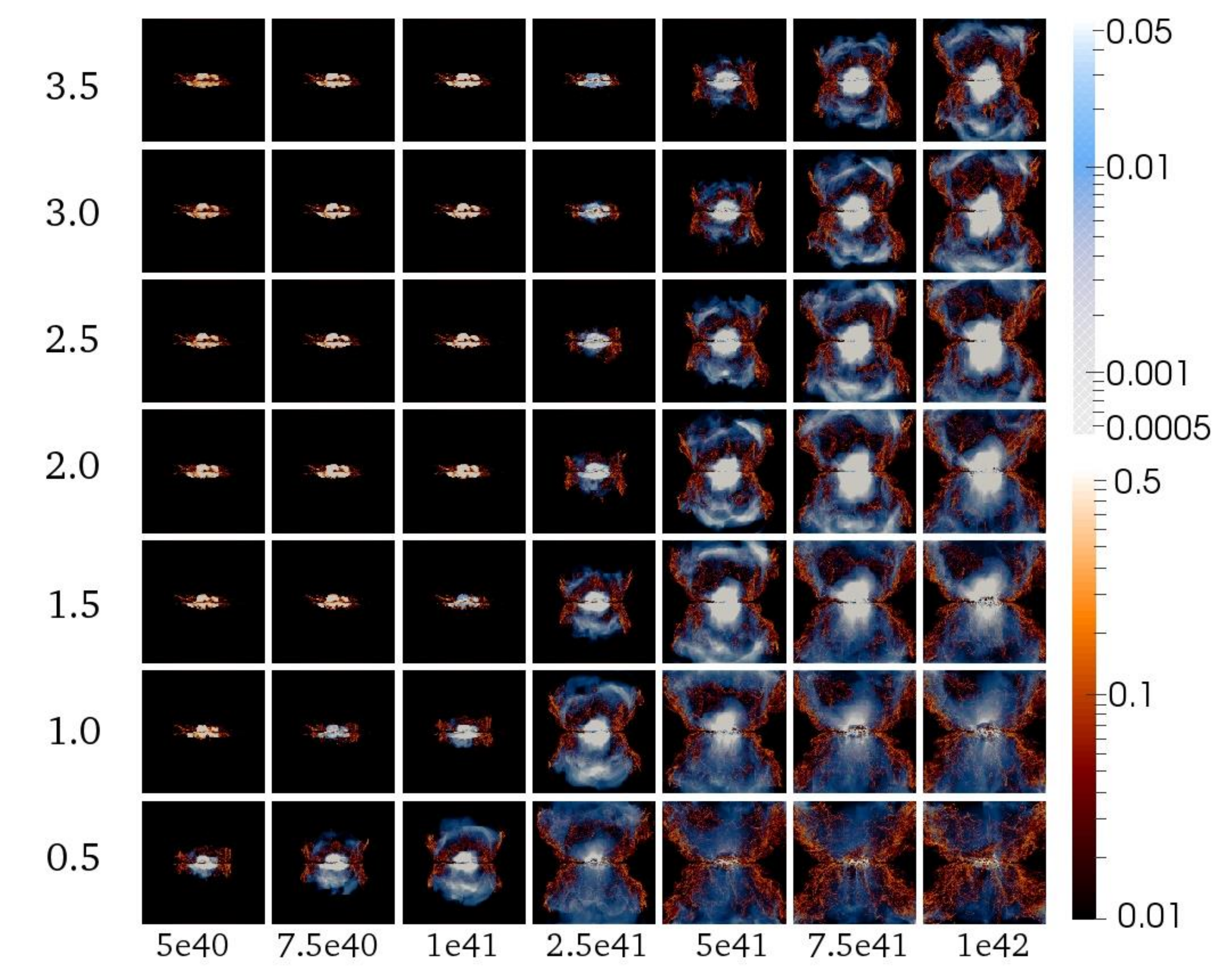}
\protect\caption{Low-resolution M1\_XXT1 models at 1.5 Myr. Models are arrayed with increasing $\dot M$ (in M$_\sun$ yr$^{-1}$) vertical and increasing $\dot E$ (in erg~s$^{-1}$) horizontal. 
Values on axes are the same as in Table \ref{tab:test:tests} and correspond to indices in model numbers. 
H$\alpha$ (red) and soft X-ray (blue) emission scaled as log(erg s$^{-1}$ cm$^{-2}$) is shown.}
\end{figure*}

Figure \ref{fig:test:128all} maps emission of H$\alpha$ and soft X-rays for the M1\_XXT1 models, viewed edge-on. Note:
\begin{enumerate}
\item Emission morphology reveals the threshold $\dot M$ and $\dot E$ for a blowout.  As expected from Eq.\ \ref{eq:body:MErel2}, larger $\dot M$ inhibits blow out but larger $\dot E$ promotes it. 
\item Soft X-rays delineate the starburst and shell of the superbubble, and fill the free wind region (Fig.~\ref{fig:test:dTHaXR34}). 
X-rays brighten with increasing $\dot M$. For low $\dot M$ but high $\dot E$ the starburst emits few X-rays. With higher $\dot M$ the hot free wind has higher mass, resulting in a significant increase in the X-ray emissivity.
\end{enumerate}

To determine which emission bands can trace a blowout we define $\Delta$ as the ratio of total emission in the lower halo ($z>85$~pc) to the disk ($z<85$~pc). Figure \ref{fig:test:WindEmis} compares $\Delta$ for different emission bands to the terminal wind speed $v_{w}$. 
Simulations with $v_{w} > 300$ \kms~have clearly experienced a blowout. Results in the blowout regime suggest the relation
\begin{equation}\label{eq:body:EmWind}
\Delta=\alpha v_{wind}^\kappa.
\end{equation}
Here $\alpha$ and $\kappa$ are constants. All bands follow the relation given in Equation \ref{eq:body:EmWind} except for the cold gas (top right panel of Figure \ref{fig:test:WindEmis}). The wind speed does not significantly affect cold gas emission, though there may be increased cold gas emission when $v_{w}>1000$ km s${}^{-1}$.
Because only two simulations (M1\_17 and M1\_27) made hard X-rays, we omit that band from our analysis. We note that the H$\alpha$ emission calculated here represents a lower bound because we do not include ionizing radiation from the stellar disk, the starburst and other sources.

\section{How Does the Cooling Function Alter Emission?}\label{sec:blowout:Emis}
We use three measures to determine how the different cooling limits affect the gas transported out of the galactic disk. We compare how T1 and T4 cooling affects the relation between $v_{w}$ and gas mass in the lower halo ($z>85$ pc), gravitationally unbound mass, and $\Delta$.

\begin{figure}[!ht]
    \centering
    \includegraphics[width=0.45\textwidth]{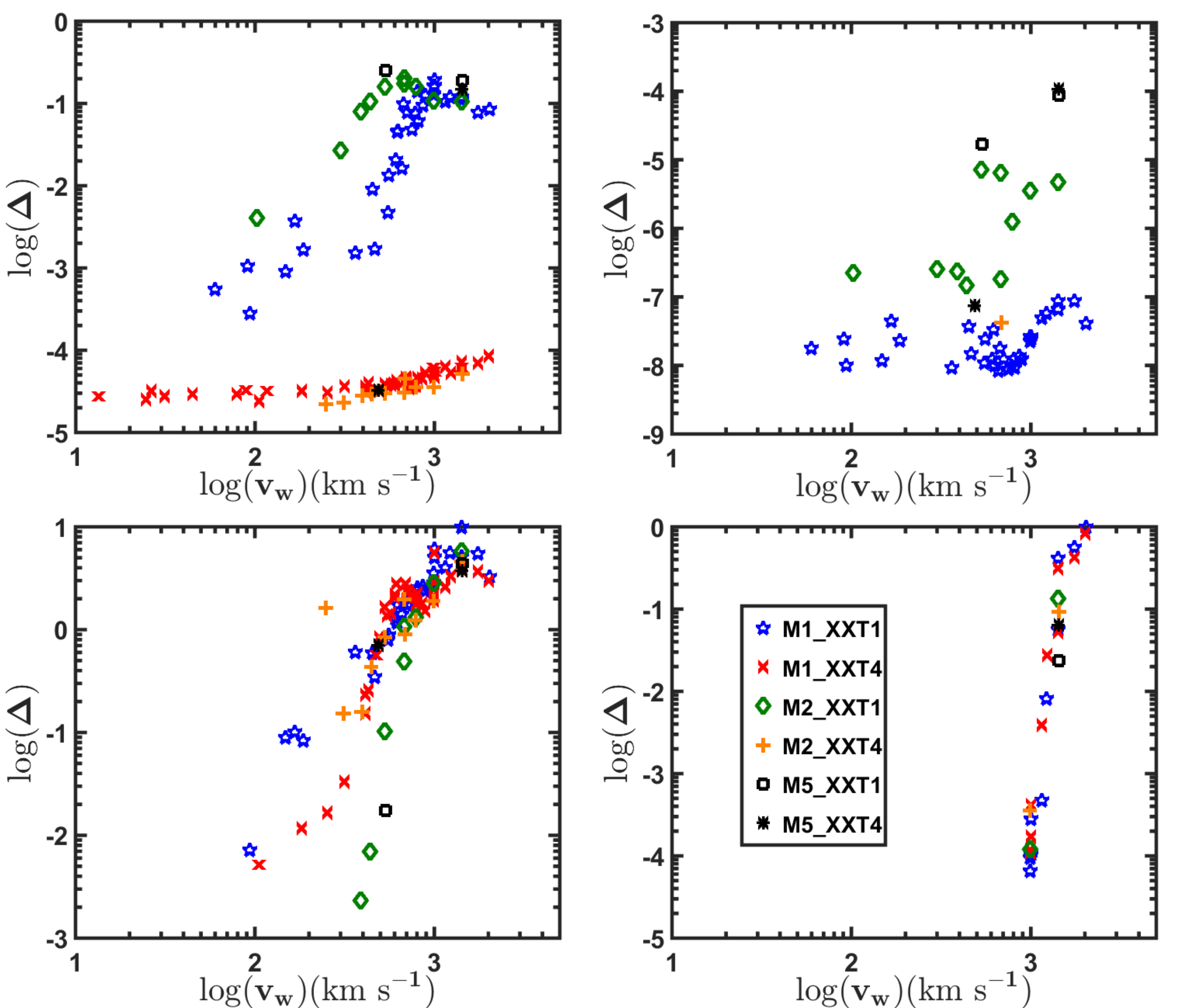}
    \caption{Total emission lower halo/disk ($\Delta$) vs. simulated wind speed at 1.5 Myr. Counterclockwise from upper right: cold gas, H$\alpha$, soft X-ray, mid X-ray.}
    \label{fig:test:WindEmis}
\end{figure}

As can be seen in Figure \ref{fig:test:WindEmis} the different cooling limits do not affect $\Delta$ for soft and mid X-rays, whereas for H$\alpha$ both $\Delta$ and $\kappa$ differ drastically between series T4 and T1. For T4 models H$\alpha$ emission in the disk is ten thousand times brighter than the lower halo, whereas for T1 models the disk is only ten times brighter. Cold gas in the lower halo ($<10^{2}$ K) emits only in T1 models. Still, lower halo emission from cold gas remains 4-8 dex below that from the disk.

\begin{figure}[!ht]
    \centering
        \includegraphics[width=0.5\textwidth]{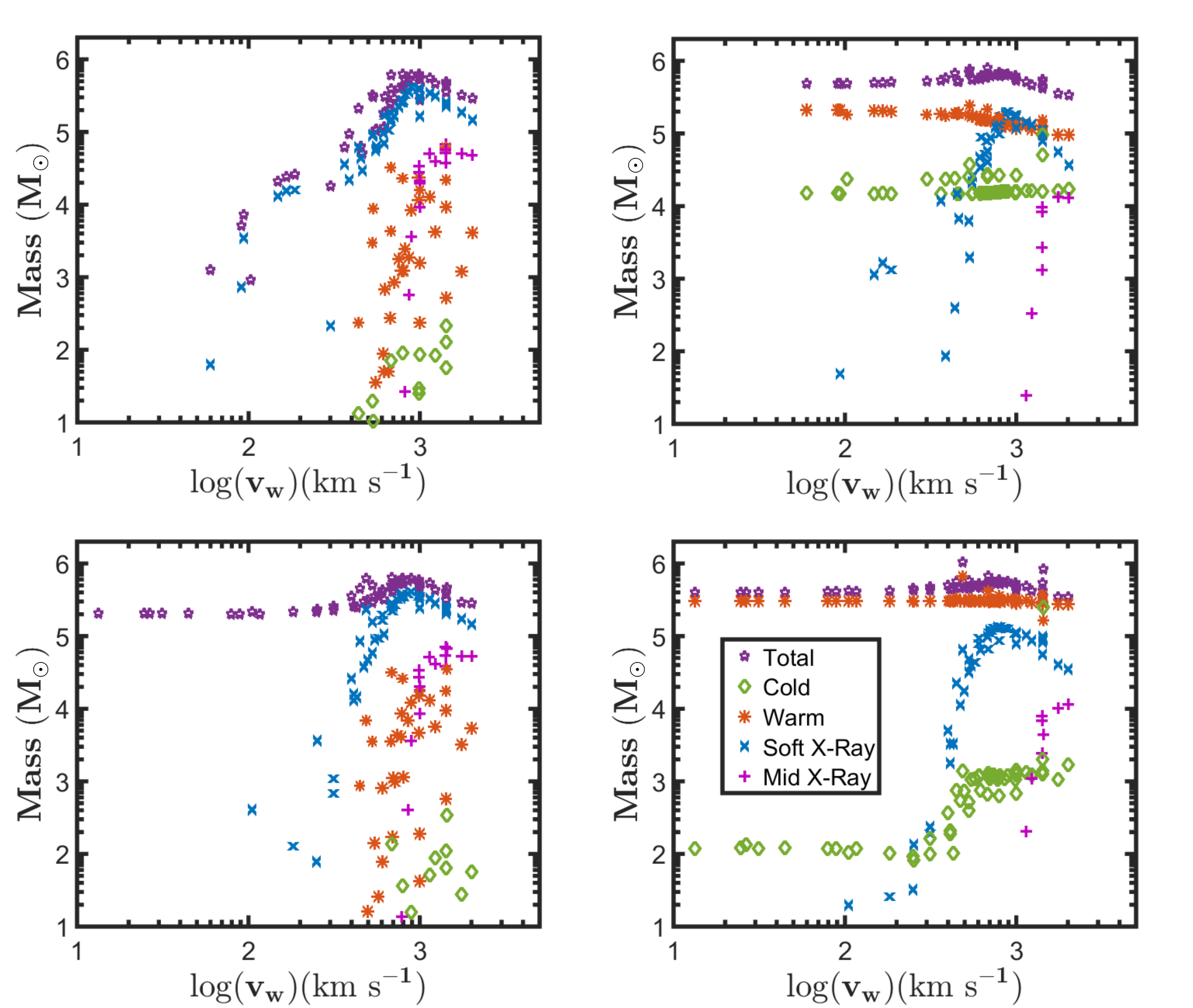}
    \caption{Gas mass vs. simulated wind speed. Graphs on the left show gas gravitationally unbound from the galaxy. On the right, gas present in the lower halo ($z>85$ pc). Graphs on the top show T1 models, on the bottom T4 models. Mass measured at 1.5 Myr.}
    \label{fig:test:WindMass}
\end{figure}

We sum the gas mass present in the lower halo ($z>85$ pc) over the central $500$~pc. We also sum the gravitationally unbound gas mass present in the disk and lower halo over the entire computational domain. Similar to \citet{StricklandStevens} we consider the gas to be gravitationally unbound if
\begin{equation}
\lvert v_z(r,z)\rvert + v_{therm}(r,z) > v_{escape}(r,z)
\end{equation}
where $\lvert v_z(r,z)\rvert$ is the bulk velocity in each cell in the vertical direction, $v_{therm}(r,z)\equiv\sqrt[]{3k_BT(r,z)/m_H}$ and $v_{escape}(r,z)$ is the escape velocity for each cell. Figure~\ref{fig:test:WindMass} plots unbound gas mass and gas mass in the lower halo vs.\ wind speed $v_{w}$ for both cooling limits. For $v_w>500$ \kms~ there is no significant difference in the unbound mass for all temperature regimes between the T4 and T1 models. Below 500 \kms~the T4 models still have $\sim 2\times 10^5$ M$_{\sun}$ of unbound mass. This mass is hot, thermally unbound, non-ballistic gas. The artificially high cooling limit of the T4 models keeps the disk gas hot and thermally unbound.

As shown in Figure~\ref{fig:test:WindMass} there is no difference in the total gas mass present in the lower halo between the T4 and T1 models. For all wind speeds, warm H$\alpha$ emitting gas dominates in T4 models but not in T1 models. 
Gas mass decreases in both at high $v_w$ because the models with highest wind speed have small $\dot M$ but large $\dot E$. Thus the wind, and by extension the lower halo, does not have as much mass.

\begin{figure}[!ht]
\centering 
\includegraphics[width=0.5\textwidth]{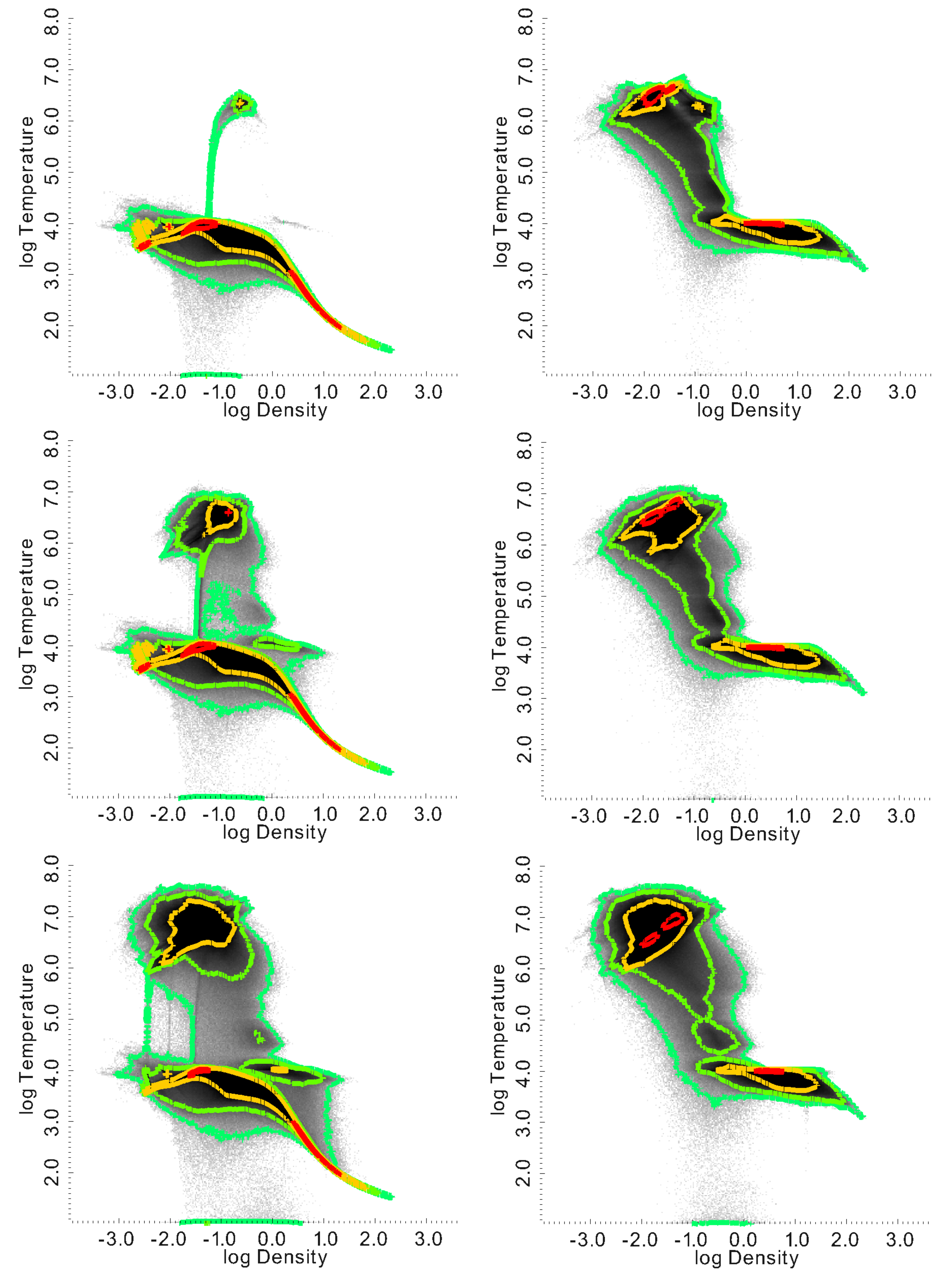}
\protect\caption{Lower halo gas mass in the temperature-density plane at 1.5 Myr. 
Left: T1 models, right: T4. Top to bottom: M2\_43, M2\_34 and M2\_25. 
Contours at $10$ (cyan), $10^{2}$ (green), $10^{3}$ (yellow), $10^{4}$ (red) M$_{\sun}$.}
\label{fig:test:Td} 
\end{figure}

Temperature-density plots in Figure \ref{fig:test:Td} demonstrate differences in model series T1 and T4: three models (M2\_43, M2\_34, M2\_25, with $\dot M$ $(2.0,1.5,1.0)$~M$_{\sun}$~yr$^{-1}$, and $\dot E$ $(1.0,2.5,5.0)\times10^{41}$~erg~s$^{-1}$) of series T1 are down the left column, and repeated for series T4 on the right. 
T4 models reproduce the H$\alpha$ ``shelf'' at $\sim10^{4}$~K of \citet{StricklandStevens} and \citet{Creasey}. 
The shelf is barely evident in T1 models.
It comprises shocked gas cooling to much lower values. 
Reduced shelf mass explains reduced H$\alpha$ gas mass in Figure \ref{fig:test:WindMass}.

Note the differing X-ray regime for model M2\_43T1 vs.\ M2\_43T4. 
In T1, cooling dominates and suppresses outflow as evidenced by an absence of hot gas in the lower halo.  
This model sits in the bottom of the intermediate regime shown in the inset in Figure \ref{fig:test:Windwind}.

\section{Embedded Filaments}\label{sec:fila}

\subsection{Expanding Bubbles}\label{sec:fila:bubbles}
Many GWs contain long optical and X-ray emitting filaments \citep{1988Natur.334...43B,1994ApJ...433...48V,1998ApJ...493..129S,1999ApJ...510..197D,1997A&A...320..378S,2002ApJ...568..689S}. 
In our simulations, filaments can form by a combination of three processes.
\begin{enumerate}
\item Limb brightening from the shocked edge of the superbubble \citep{2002ApJ...576..745C}.
\item Disruption of a cool dense cloud by the supersonic wind \citep{2002ApJ...576..745C,CooperII}.
\item Merging bubbles that rise from the starburst region \citep{2006ApJ...653.1266J,2013MNRAS.430.3235M}.
\end{enumerate}

Limb brightened filaments appear in Figures \ref{fig:test:128all} and \ref{fig:test:dTHaXR27} at the edge of the shocked region; they are broad ($100-200$ pc) without well defined boundaries. These filaments have no significant vertical motion because they represent the edge of the wind region. Embedded in these regions may be smaller filaments formed through processes 2 and 3 as discussed below.

Cold dense clouds are overrun by the supersonic hot wind, which exerts a ram pressure on the cloud, disrupting it, stripping off material and elongating it into a filament. Examples of disrupted clouds can be seen in the density plots in Figures \ref{fig:test:dTHaXR34} and \ref{fig:test:dTHaXR27}. While these disrupted clouds are present in our simulations, we cannot fully resolve them because that would require a resolution of $<0.1$ pc (see \citet{CooperII}) compared to our maximum of $2$ pc.

Due to the inhomogeneities in the starburst multiple bubbles form which sweep up and squeeze the ISM. With continued expansion, the shells merge to coalesce the gas into thin ($<~50$ pc) filaments. In our models, many of these filaments emit little H$\alpha$ before dispersing within a Myr by shock heating and ablation, or disrupting by Kelvin-Helmholtz instabilities.

\begin{figure}[!ht]
\centering
\includegraphics[width=0.25\textwidth]{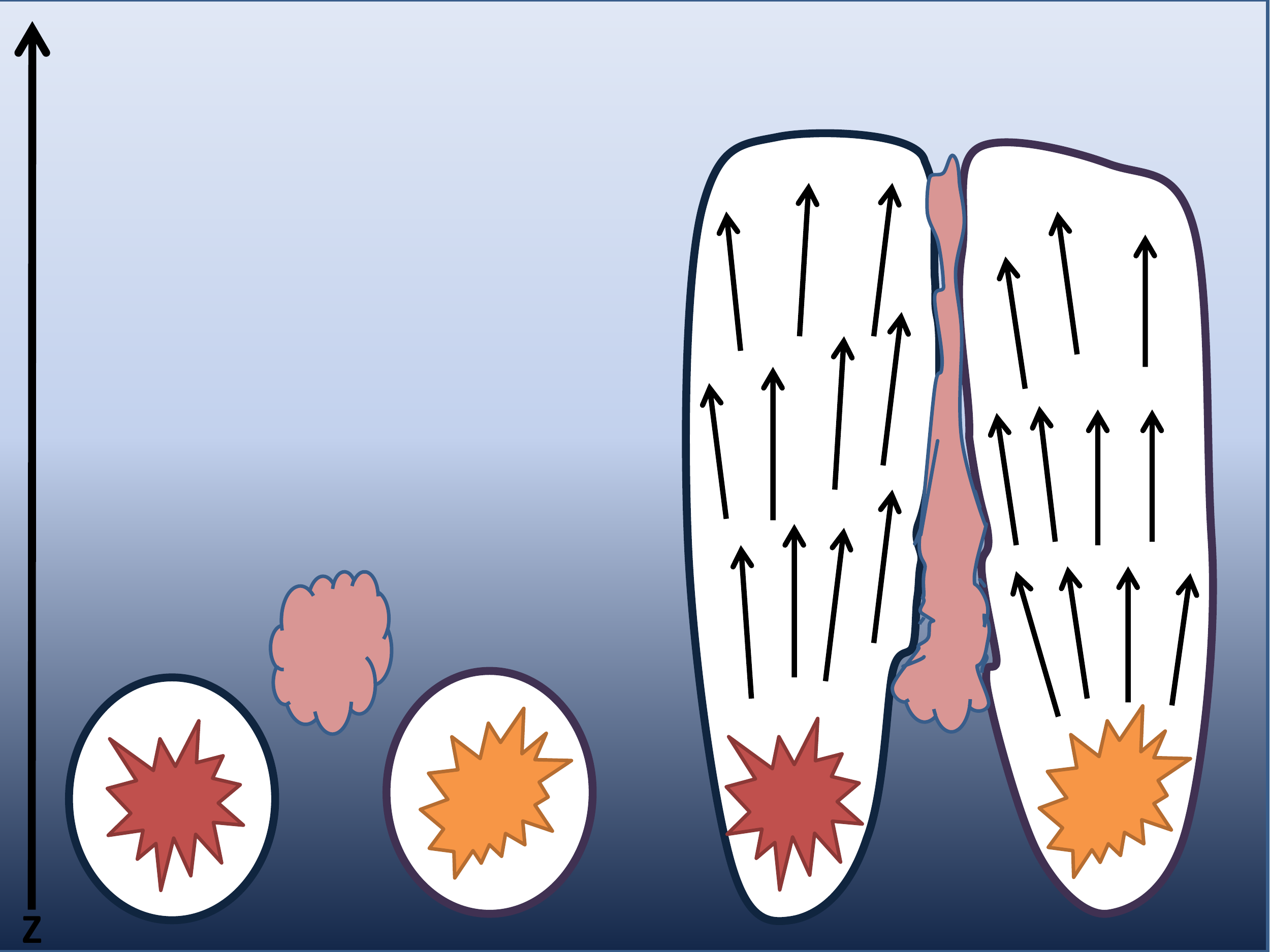}
\protect\caption{Cartoon of two merging superbubbles viewed side-on, combining filament formation scenarios 2 and 3.
Their contact forms a filament from ISM swept up and compressed by the wind. 
To persist, this filament must be anchored to a mass loading source to continuously replenish its shocked, dense gas. 
}
\label{fig:fila:cartoon} 
\end{figure}

A few filaments persist when a cool dense cloud is present along the bubble merger interface. The additional mass allows the filament to persist longer before disrupting entirely. If the filament is anchored to a mass loading site within the starburst, the dense gas in the filament can be replenished continuously allowing the filament to survive for more than a Myr and stretch to $>100$~pc. Figure \ref{fig:fila:cartoon} sketches this last scenario which is a combination of processes 2 and 3 above.

Figures \ref{fig:test:DV34} and \ref{fig:test:DV27}, which compare models M5\_34T1 and M5\_27T1 respectively, show examples of filaments forming through a combination of cloud disruption and merging bubbles. These filaments are embedded in a GW of $400<v<2000$ \kms. The densest material has a velocity of $\lesssim50$ \kms~whereas ablated material $200<v<500$ \kms. Thus the dense cores of the filaments are hardly moving with respect to the disk. The wind flows by, ablating and collimating the filaments.
The velocity gradient of its ablata resembles the homologous $v(r)\propto r$  velocity gradient mapped in NGC 3079, although velocities are lower than the 1500 \kms~observed \citep{2001ApJ...555..338C,2002ApJ...576..745C}. 
 
The strength of the GW determines how filaments evolve. We note two interesting cases outlined below. 

\subsection{Mass Anchors}\label{sec:fila:anchor}
\begin{figure}[!ht]
\centering 
\includegraphics[width=0.5\textwidth]{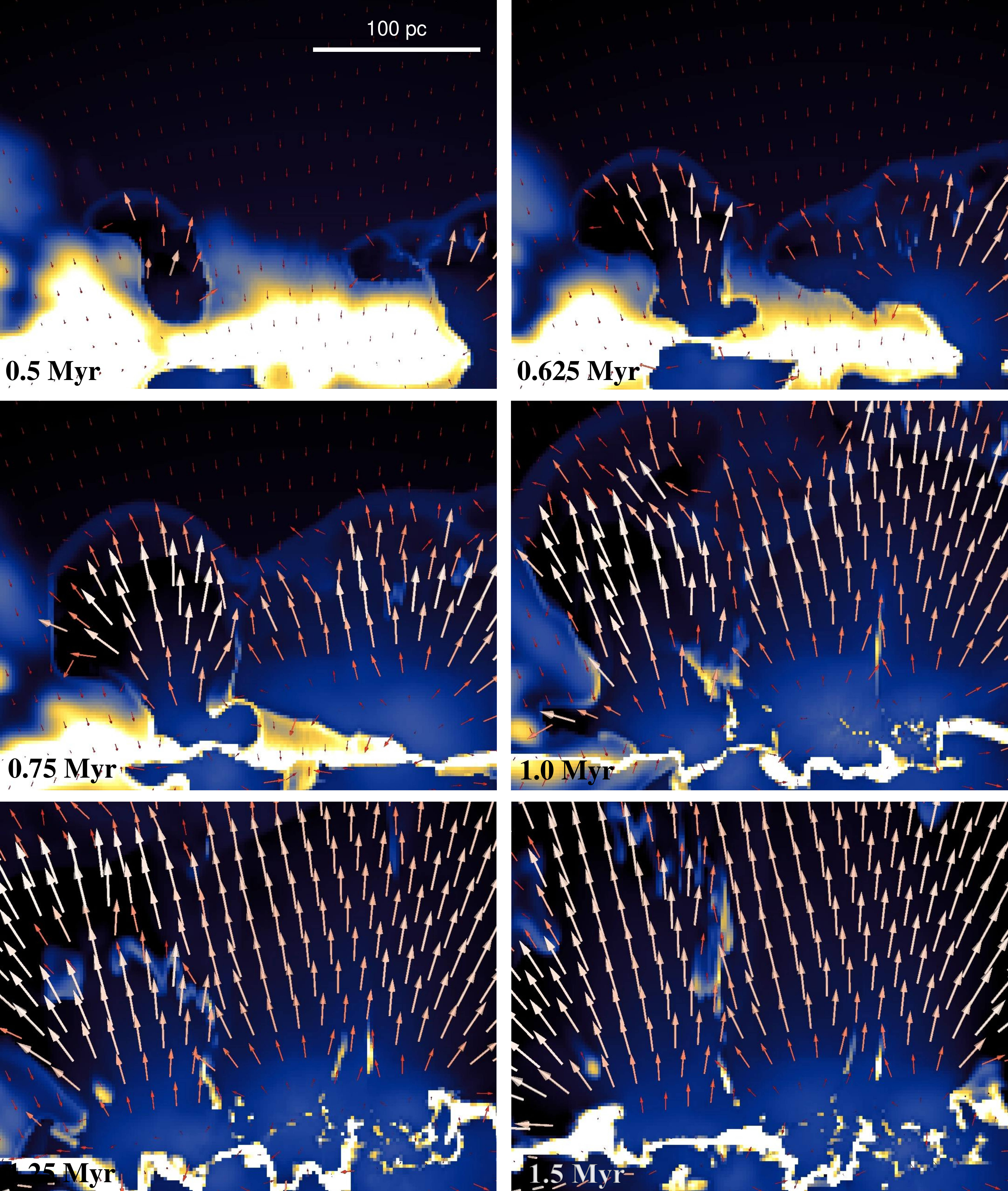}
\protect\caption{Close-up of the filament forming region delineated in Figure \ref{fig:test:dTHaXR34} (model M5\_34T1, bottom right panel). 
The starburst covers the bottom third of each image.  
Red velocity vectors are $v_{w}\approx 20$ \kms~and white $v_{w}\approx 500$ \kms. The filament is forming just left of center where the two bubbles are merging.}
\label{fig:test:DV34} 
\end{figure}

Model M5\_34T (Fig.\ \ref{fig:test:DV34}) has sufficient energy to form a GW, but the wind does not entirely disrupt all filaments. As shown in Figure \ref{fig:test:DV34} two distinct bubbles emerge from the central starburst. Their boundaries merge to form a dense filament that stretches $>100$~pc back to anchor on the starburst reservoir. The 540 \kms~wind ablates mass off the reservoir, pushes it into the filament, and by 1.5 Myr has extended the filament $>400$ pc above the disk plane where it is drifting along at only $50-100$ \kms. Due to continual mass loading at the base of the filament it stays anchored, allowing it to persist and grow. At some point the filament should disrupt entirely due to either Kelvin-Helmholtz instabilities or heating and evaporation. But our resolution is insufficient to maximize filament survival time \citep[see][]{CooperII}.

\subsection{Filament Lift Off}\label{sec:fila:liftoff}
\begin{figure}[!ht]
\centering
\includegraphics[width=0.5\textwidth]{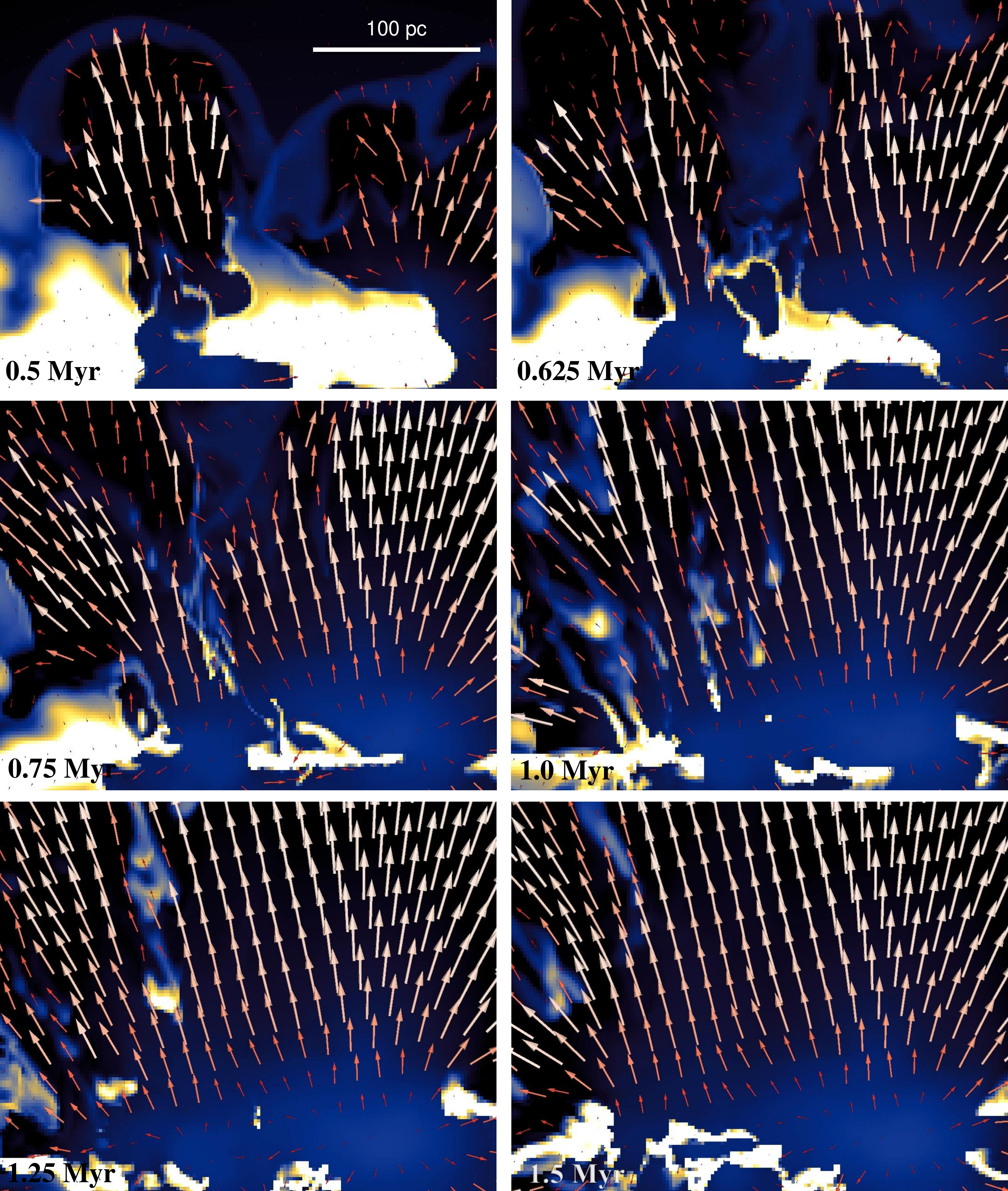}
\protect\caption{Same as Figure \ref{fig:test:DV34}, for model M5\_27T1. Velocity vectors are color-coded, ranging from 20 \kms~to 1500 \kms.}
\label{fig:test:DV27} 
\end{figure}

In model M5\_27T1 (Fig.\ \ref{fig:test:DV27}) the filament again forms along the bubble contact. 
But now, after 1 Myr it detaches from the disk reservoir and lofts into the free-flowing wind of the now merged bubbles. 
This filament differs from its slow counterpart model M5\_34T1; it has a larger cross section to the impinging wind, so it fragments more due to Kelvin-Helmholtz instabilities.
The surrounding wind flows at $1420$ \kms~while the filament moves at $0-50$ \kms~before lift off but attains $200-500$ \kms~thereafter. This filament would be analogous to the disrupted clouds studied by \citet{CooperII}.

\section{Synthetic Absorption Lines}\label{sec:lines}
We showed in \S\ref{sec:blowout:Emis} that cold gas emission entrained in the GW is 4-5 dex fainter than the galactic disk.
A more sensitive probe of this gas may be its absorption of background starburst continuum. 
We synthesize absorption lines for three temperature regimes, denoted ``molecular'', ``warm'', and ``soft X-ray'', with temperature ranges given in Table \ref{tab:body:TempEmis}. 
A trivial line source function suffices for kinematical signatures of the three temperature regimes. 

Absorption spectra are derived by integrating optical depth in $N$ cells along the column viewed perpendicular to the disk 
\begin{equation}
\tau(v_{ch})=\sum_{i}^{N}\tau_{i}(v_{ch}).
\end{equation}
The velocity channels have a resolution of 10 \kms~and range from -1800 \kms~to 200 \kms. 

\begin{figure}[!ht]
\centering
\includegraphics[width=0.47\textwidth]{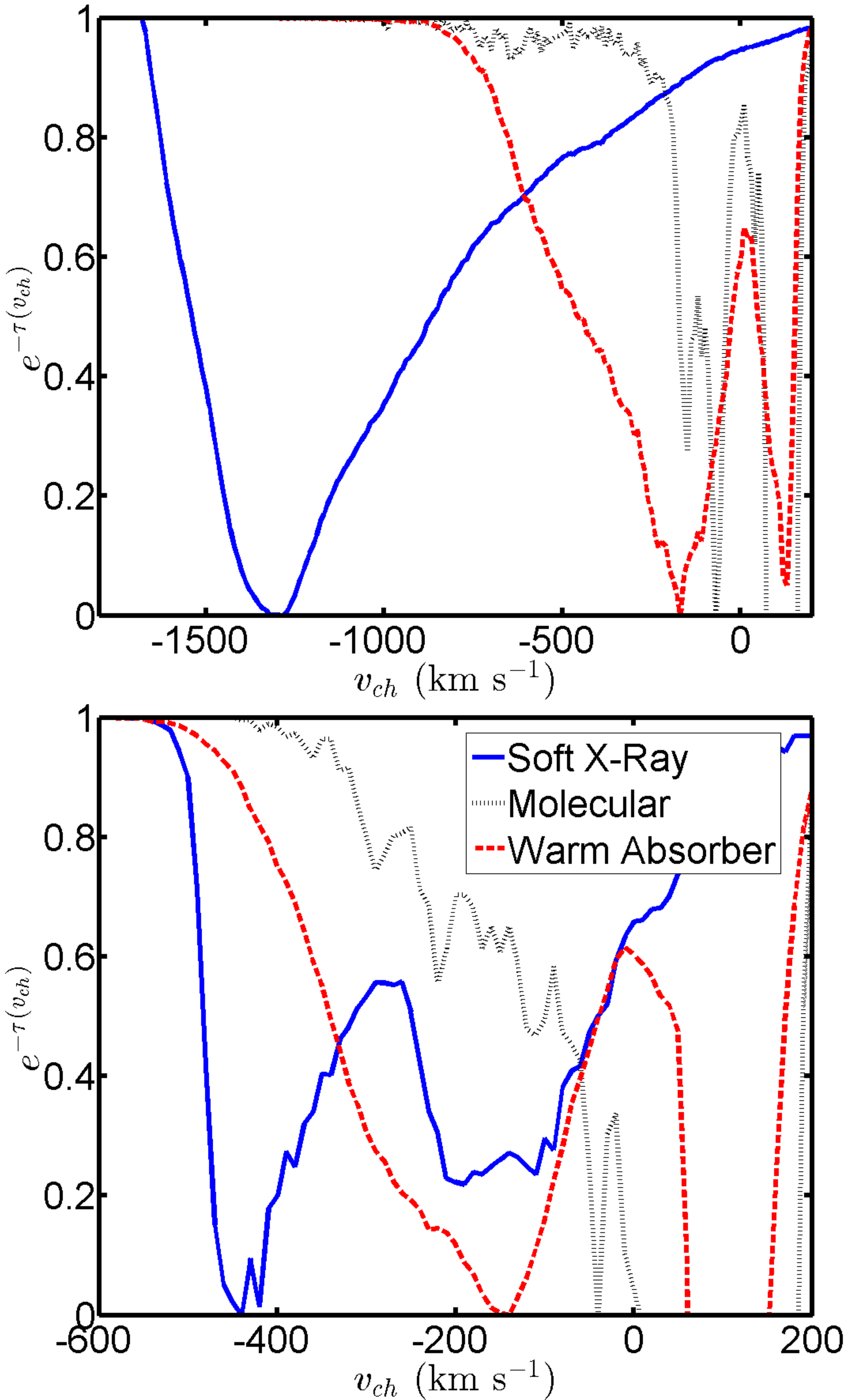}
\protect\caption{Synthetic absorption line profiles for model M5\_27T1 (top) and M5\_34T1 (bottom). 
Absorptions are calculated for ``soft X-ray'', ``molecular'', and ``warm'' gas.
Vertical normalization is arbitrary.\label{fig:lines:3427-1}}
\end{figure}

Absorption profiles are shown in Figure \ref{fig:lines:3427-1} for models M5\_27T1 (top panel) and M5\_34T1 (bottom).
The ``soft X-ray'' line shows the structure of the hot free-wind inside the expanding bubble. 
The velocity at maximum absorption is the average speed of the free wind.
The long tail of the profile back toward galaxy systemic velocity, especially prominent in model M5\_27T1,  reveals gas flowing radially at the average speed of the free wind but not entirely along our line of sight.
Model M5\_34T1 shows two spikes in this absorption profile. 
The faster corresponds to the free wind inside the expanding bubble, the slower to absorption in the bubble shell. 
This shell has left the computational grid in model M5\_27T1. 

The ``warm'' line traces filaments and clouds caught in the gas but moving much slower, so maximum extinction is at a much lower velocity. 
The long tail of this profile traces ablata accelerating off the filaments.

The ``molecular'' line shows a similar tail, although that absorption is more varied because multiple clouds contribute. In both the ``warm'' and ``molecular'' profiles there is absorption for positive velocities. This results from dense clouds perturbed by the shock initially found at the edge of the lower halo which have begun to fall towards the galactic plane. 
For arbitrary absorption in the neutral medium, we would expect an acceleration tail similar to that in the warm and molecular lines.

The asymmetric ``warm'' and ``molecular'' absorption line profiles are similar to observed \ion{Si}{2}, \ion{Si}{3}, \ion{O}{1}, \ion{C}{2} \citep[see][Fig. 11, especially KISSR 242 and KISSR 1578]{2013ApJ...765..118W}, and Ly$\alpha$ \citep[see][Figs.\ 5 and 6]{2012ApJ...751...51J} profiles in starburst galaxies. 
The shape also matches analytical predictions \citep{2015ApJ...801...43S}.

Positive absorption features shown in Figure \ref{fig:lines:3427-1} result from clouds initially at the edge of the lower halo, but not directly above the starburst. They were perturbed by the shock from the starburst but not blown out by it and have begun to fall towards the disk.

\section{Discussion}\label{sec:discuss} 

\subsection{Blowout Conditions}\label{sec:discuss:blowout} 
In Eq.\ \ref{eq:body:MErel}, $\xi$ measures the fraction of $\dot E$ converted into wind kinetic energy. \citet{2009ApJ...698..693F} calculated  $\xi=0.45$ whereas our models found $0.67$. 
The difference between our value of $0.67$ and the analytic $0.45$ can be attributed to two causes:
\begin{enumerate}
\item More starburst $\dot E$ goes into the kinetic energy of the wind because less energy is being expended to push through the inhomogeneous ISM.
\item More loaded mass $(\dot{M})$ ends up in filaments and is not accelerated to the terminal wind speed (\S \ref{sec:fila}), so is not draining starburst energy.
\end{enumerate}
Our simulations cannot establish which of these dominates. The specific value of $\xi$ may depend on parameters such as gas surface density \citep{Creasey} and ambient ISM pressure \citep{1988ApJ...324..776M}.

When considering the analytic wind speed ($v_A$ from Eqn. \ref{eq:body:MErel2}), there is a transition region ranging from escape velocity ($v_e$) to $1.5v_e$ where a wind can form but its evolution is set by cooling and resolution (Fig.\ \ref{fig:test:Windwind} inset). 
Within this region our T4 models have faster winds while the corresponding T1 models sometimes have no wind.
This difference arises because our T1 models lose more energy to cooling.
Above the transition, cooling has no effect on the kinematics of a blowout, in agreement with \citet{2009ApJ...698..693F}; moreover, increased resolution does not alter the measured wind speed. 

Across the transition, higher resolution models form a GW at low $v_A$ but the corresponding lower resolution models do not; e.g.\ both M2\_33T4 and M2\_33T1 formed a wind but the M1\_33T4 and M1\_33T1 models did not. 
But at a lower $v_A$ the M2\_43T4 model formed a wind while the M2\_43T1, M1\_43T4, and M1\_43T1 did not despite having the same calculated $v_A$. 
This explains the absence of hot gas in the upper left panel of Figure \ref{fig:test:Td}.
Higher resolution models form more filaments and dense cores, which decreases overall cooling efficiency. 
Lower resolution models over-estimate cooling losses.
We did not run mid- or high-resolution models below the escape velocity, so cannot say if a starburst will blow out if the analytic wind speed is below $v_e$.

While our analysis was done at 1.5 Myr, our low resolution models ran to 4 Myr. 
If a blowout is absent at 1.5 Myr, it is also absent at 4 Myr. 
We conclude that an instantaneous starburst with constant mass and energy injection will reach terminal wind speed before 1.5 Myr.

\subsection{Effect of the Radiative Cooling Limit}\label{sec:discuss:cooling}
Numerical studies of starbursts with radiative cooling have focused on the warm wind plasma at $T>10^4$~K \citep{StricklandStevens,SutherlandBicknell,CooperI,2011ApJ...740...75W,Creasey,2013MNRAS.430.3235M,Williamson11062014}, with a few addressing $100$~K gas \citep{2006ApJ...653.1266J,2009ApJ...698..693F,Hill}. 

We confirm \citet{2009ApJ...698..693F} that T4 cooling suffices if one is interested only in kinematics and when $v_A > 1.5v_e$; GW formation depends only on mechanical luminosity of the starburst and associated mass loading from the stellar winds. 
Histograms in Figure \ref{fig:test:Td} for T4 cooling resemble Figure 3 of \citet{Creasey}, showing a ``shelf" of H$\alpha$ emission at $10^4$ K. 
But Figure \ref{fig:test:Td} with T1 cooling shows that the \citet{Creasey} ``shelf'' is an artifact of T4 cooling and we showed in \S\ref{sec:blowout:Emis} that GW composition changes significantly.
The GW is no longer dominated by H$\alpha$ emitting gas, and, in agreement with \citet{2013Natur.499..450B}, is dominated by neutral, molecular, and X-ray emitting gas.

The ratio of X-ray emission in the lower halo to that in the disk is unaffected by T1 cooling, but there is a change of 1-3 dex in the ratio of H$\alpha$ emission.

\subsection{Resolution}\label{sec:discuss:res}
To examine the effect of resolution we ran our MX\_34 and MX\_27 models at three different resolutions, and compared the wind velocities, lower halo mass, and unbound mass in the different temperature regimes. As noted in Section \ref{sec:problem:frac} the same initial density distribution was used for all models and was coarsened for the lower resolution models. Additionally our M5\_27 and M2\_27 models use the same parameters and resolutions as model numbers M01 and M04, respectively from \citet{CooperI}.

For our MX\_34 and MX\_27 models we find no difference in wind velocity within the uncertainty once a steady state wind had formed after 1.5 Myr. For all MX\_34 models $v_{w} \approx 550$ \kms~and for all MX\_27 models $v_{w} \approx 1420$ \kms. As shown in Figure \ref{fig:test:Windwind} for $v_{w} > 500$ \kms~the relation given in Equation \ref{eq:body:MErel} holds irrespective of resolution. Thus the wind kinematics of a sufficiently powerful starburst are not affected by numerical resolution. But note, when $v_w < 500$ \kms~(see Figure \ref{fig:test:Windwind} insert) the formation of a wind depends on the resolution. Lower resolution models may experience enhanced cooling due to greater average density from unresolved features. Thus for models on the edge of a blowout increased resolution is important for determining if a GW will form.

\begin{figure}[!ht]
\centering
\includegraphics[width=0.47\textwidth]{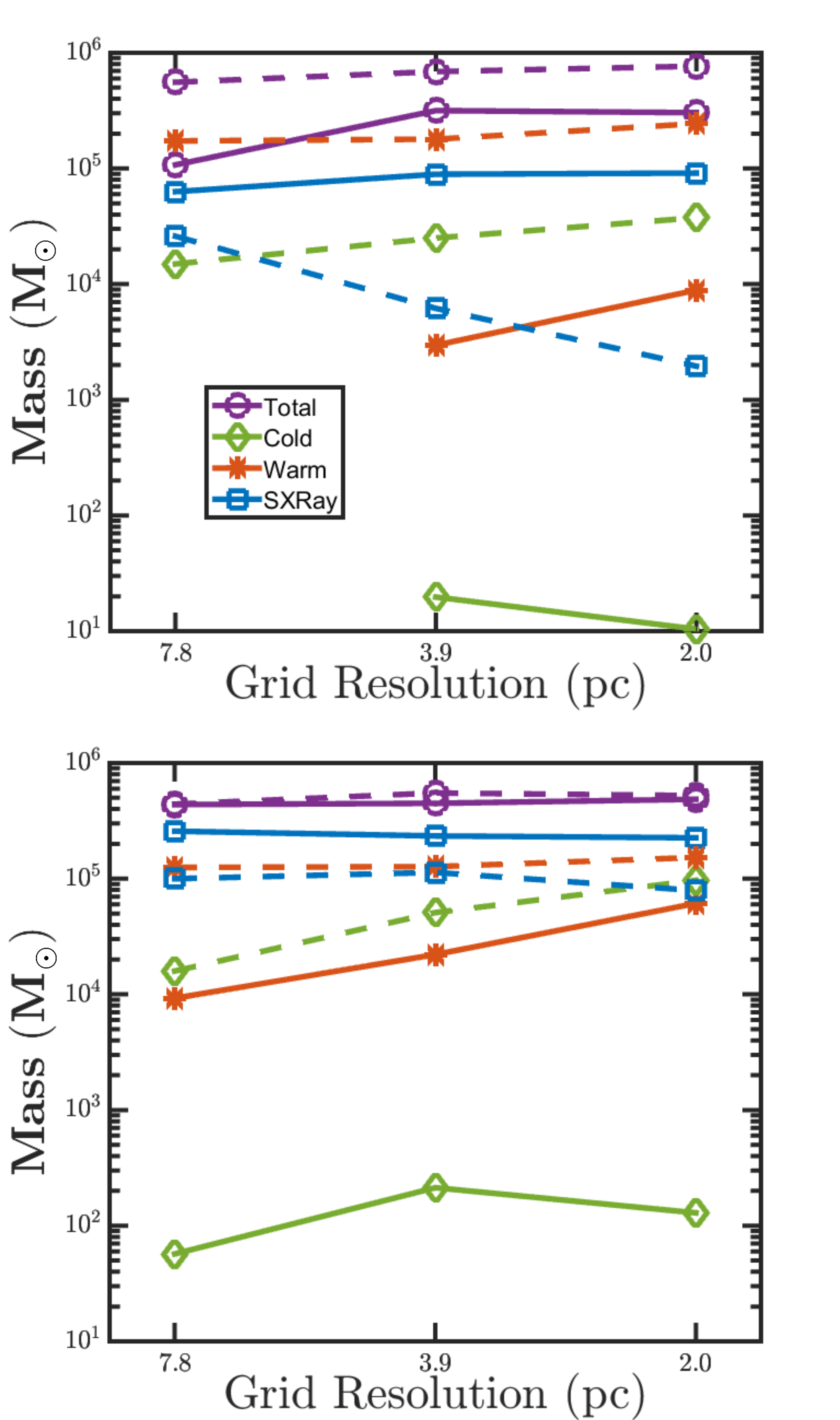}
\protect\caption{Lower halo and unbound gas mass at different grid resolutions. Solid lines indicate unbound mass, dashed lines indicate lower halo mass. Top MX\_34T1 models, bottom MX\_27T1 models.\label{fig:Res}}
\end{figure}

As shown in Figure \ref{fig:Res}, similar to the wind speed noted above, increased resolution does not significantly change the total unbound and lower halo mass, with the exception of the M1\_34 model. The M1\_34 model is just above the limit of $v_w < 500$ \kms~where resolution begins to affect the kinematics. This is evident as a slight decrease in the total unbound mass at the lowest resolution. The unbound mass of soft X-ray gas is not affected by resolution for both sets of models, but for our MX\_34 models there is marked decrease in soft X-ray gas mass in the lower halo. This is due to the increased resolution of bow shocks and hot envelopes surrounding filaments which decreases the amount of mass in that temperature regime. This effect is not seen in the MX\_27 models because the superbubble has expanded to fill the entire lower halo volume. Here the mass contribution of bow shocks and hot envelopes surrounding filaments is not as significant. Related to this is an increase in unbound, warm, H$\alpha$ emitting gas from ablata off of ballistic filaments. This corresponds to increased cold gas in the lower halo as higher resolution models form more well defined filaments containing cold gas.

\subsection{Filaments}\label{sec:discuss:filaments}
Section \ref{sec:fila} listed three origins of emitting filaments in our simulations.
The longest filaments are from limb brightening and trace the bottom half of the expanding superbubble. 
Filaments from disrupting cold clouds or merging bubbles are thinner and shorter. 
Filaments from merged bubbles have higher densities and more optical emission \citep[see][]{2006ApJ...653.1266J}, thus do not just arise from projection like limb brightened filaments.

Our model resolution sufficed only to outline filaments. 
As \citet{CooperII} note, better resolution of filaments merely increases gas fragmentation and number of cloudlets, but does not change their kinematics. While Cooper et al. did not include thermal conduction in their simulations they noted that it should decrease cloud fragmentation by suppressing Kelvin-Helmholtz instabilities \citep{2000Ap&SS.272..189V,2007A&A...472..141V}. They concluded that it should increase cloud survival time despite an increase in mass lost due to evaporation.
They found that radiative cooling contributes to filament survival. 
They used MAPPINGS III \citep[based on][]{1993ApJS...88..253S}, which only extends down to $10^4$ K. 
If cooling below $10^4$ K is allowed, more cloudlets would survive to transport cold gas into the galactic halo. While \citet{CooperII} considered the disruption of a cloud embedded in a GW, an interesting extension of their work would be to model a cloud anchored to a mass loading region as explained in \S\ref{sec:fila:anchor}.

We find that the cold mass blown into the lower halo does not depend on starburst strength.
\citet{2013MNRAS.430.3235M} showed that more dense packing of young clusters within a starburst forms more filaments. 
This may be due to more contact between expanding bubbles.
We show that filaments form along contacts and persist when attached to a mass loading anchor. 
There is a higher probability of contacts and anchors with many star forming complexes within the starburst.
Most cold mass blown into the lower halo by the GW is filamentary, only a bit remains in dense clouds that are not disrupted.

It is interesting that starburst luminosity does not alter the cold mass swept up by the GW. 
This may indicate \citep{2013MNRAS.430.3235M} that the cold mass blown into the lower halo is set by the initial distribution of dense ISM clouds and the density of new star clusters within the starburst.

\section{Conclusions}\label{sec:end}
Our two series of 3D simulations explore how a wide range of mechanical luminosity and mass loading of a nuclear starburst affects GW formation in an M82 sized galaxy.
We also compare how gas cooling to $10^4$ K (T4) vs.\ 10 K (T1) affects outflow emission and loaded mass. 
We conclude that:
\begin{enumerate}
\item The threshold for a blowout is when $v_{A}>1.5v_e$ with $v_A$ defined by Equation \ref{eq:body:MErel2}. Below this limit the possibility of a blowout depends on the cooling and grid resolution used. Above this limit cooling and grid resolution do not affect wind kinematics.
\item For T4 cooling, most lower halo gas is in the warm regime corresponding to peak H$\alpha$ emission. 
But for T1 cooling, lower halo mass is predominantly neutral, cold and X-ray emitting, not warm H$\alpha$ emitting gas.
\item Soft and mid X-rays from edge-on starburst galaxies trace the strength of a GW because the ratio halo/disk emission correlates with GW terminal speed.
\item Emission from cold gas in the lower halo is 4-8 dex fainter than from cold gas in the disk.
\item The mass of cold gas blown into the lower halo does not depend on starburst strength. It may depend on the ISM initial state and the number of star-forming complexes \citep{2013MNRAS.430.3235M}.
\item Bright optical filaments form in 3 ways. 
Observed filaments can be any combination of:
\begin{enumerate}
\item Limb brightened, shocked edge of the superbubble.
\item A cool dense cloud ablated by the wind.
\item Merged bubbles that rise from the starburst.
\end{enumerate}
\item Filaments move much slower than the wind. Filaments embedded in a GW of $400<v<2000$ \kms~attain $\lesssim50$ \kms~for the densest material and $200<v<500$ \kms~for ablata.
\item The densest filaments form molecular and ``warm" absorption line profiles that are asymmetric with long tails to higher velocities from accelerating ablata.  They resemble those observed in starbursts.
\item T1 cooling of a sufficiently powerful nuclear starburst does not change GW kinematics, confirming \citet{2009ApJ...698..693F}
\item Absorption lines from warm and cold dense gas can be used to infer the terminal velocity of the hot diffuse wind. A measurement of the velocity of the GW can be used to infer the size of the generating starburst, using Eqs. \ref{eq:body:MErel2} and \ref{eq:body:MErel}, even when the starburst cannot be measured directly.
\end{enumerate}

\acknowledgments{}
NASA Herschel grants NHSC-OT-1-1436036 and NC Space Grant supported this work.

\appendix\label{sec:app}
\subsection{Radiative Cooling in Athena}\label{sec:app:cool}
This public code handles radiative cooling by adding an external source term given by Equation \ref{eq:heatingandcooling} to the energy equation within the CTU integrator. 
Substantial $T$ and pressure gradients in our simulations require modification to improve the accuracy of the cooling step by sub-cycling a 2/3rd order adaptive step-size integrator \citep{Bogacki1989321}, as follows.
For each cell at each time step, $\Delta T$ is calculated using a single pass through the Bogacki-Shampine method. 
If the difference between the 2nd and 3rd order results exceeds 10\% or if the method returns a non-physical result then $\Delta T$ for the cell is recalculated using an adaptive step subroutine. 
Otherwise, we keep the result from the first pass.

As the cooling step ends we check if the calculated $\Delta T$ deviates the cell from its radiative equilibrium $T$ at its current density. 
We also impose a 10 K floor to ensure a physical result.

\subsection{Kinetic Flux Vector Splitting}\label{sec:app:kfvs}
We add a backup way to calculate fluxes for the 1-5 cells (out of $6\times N^{3}$ flux calculations) in a single time step where the normal calculation using the hllc solver returned a non-physical result. 
The fall-back algorithm, Kinetic Flux Vector Splitting \citep{Mandal1994447}, solves the collisionless Boltzmann equation. 
While more diffusive, it stabilizes at rarely encountered, extreme gradients. 
Because very few cells are affected, the overall diffusiveness of the code does not change.

\subsection{Integrator Modifications}\label{sec:app:int}
Our simulations encountered a few cases at the cell walls where the high-order interpolator returned negative densities.
To set a floor on density, we use a first-order (piece-wise constant) interpolation over density at those failures.

\medskip
\bibliography{./generalbib}

\begin{thebibliography}{}
\expandafter\ifx\csname natexlab\endcsname\relax\def\natexlab#1{#1}\fi

\bibitem[{{Aguirre} {et~al.}(2001){Aguirre}, {Hernquist}, {Schaye}, {Weinberg},
  {Katz}, \& {Gardner}}]{2001ApJ...560..599A}
{Aguirre}, A., {Hernquist}, L., {Schaye}, J., {et~al.} 2001, \apj, 560, 599

\bibitem[{{Barker} {et~al.}(2008){Barker}, {de Grijs}, \&
  {Cervi{\~n}o}}]{2008A&A...484..711B}
{Barker}, S., {de Grijs}, R., \& {Cervi{\~n}o}, M. 2008, \aap, 484, 711

\bibitem[{{Bland} \& {Tully}(1988)}]{1988Natur.334...43B}
{Bland}, J., \& {Tully}, B. 1988, \nat, 334, 43

\bibitem[{Bogacki \& Shampine(1989)}]{Bogacki1989321}
Bogacki, P., \& Shampine, L. 1989, Applied Mathematics Letters, 2, 321

\bibitem[{{Bolatto} {et~al.}(2013){Bolatto}, {Warren}, {Leroy}, {Walter},
  {Veilleux}, {Ostriker}, {Ott}, {Zwaan}, {Fisher}, {Weiss}, {Rosolowsky}, \&
  {Hodge}}]{2013Natur.499..450B}
{Bolatto}, A.~D., {Warren}, S.~R., {Leroy}, A.~K., {et~al.} 2013, \nat, 499,
  450

\bibitem[{{Cecil} {et~al.}(2002){Cecil}, {Bland-Hawthorn}, \&
  {Veilleux}}]{2002ApJ...576..745C}
{Cecil}, G., {Bland-Hawthorn}, J., \& {Veilleux}, S. 2002, \apj, 576, 745

\bibitem[{{Cecil} {et~al.}(2001){Cecil}, {Bland-Hawthorn}, {Veilleux}, \&
  {Filippenko}}]{2001ApJ...555..338C}
{Cecil}, G., {Bland-Hawthorn}, J., {Veilleux}, S., \& {Filippenko}, A.~V. 2001,
  \apj, 555, 338

\bibitem[{{Chevalier} \& {Clegg}(1985)}]{1985Natur.317...44C}
{Chevalier}, R.~A., \& {Clegg}, A.~W. 1985, \nat, 317, 44

\bibitem[{{Cooper} {et~al.}(2008){Cooper}, {Bicknell}, {Sutherland}, \&
  {Bland-Hawthorn}}]{CooperI}
{Cooper}, J.~L., {Bicknell}, G.~V., {Sutherland}, R.~S., \& {Bland-Hawthorn},
  J. 2008, ApJ, 674, 157

\bibitem[{{Cooper} {et~al.}(2009){Cooper}, {Bicknell}, {Sutherland}, \&
  {Bland-Hawthorn}}]{CooperII}
---. 2009, ApJ, 703, 330

\bibitem[{{Creasey} {et~al.}(2013){Creasey}, {Theuns}, \& {Bower}}]{Creasey}
{Creasey}, P., {Theuns}, T., \& {Bower}, R.~G. 2013, \mnras, 429, 1922

\bibitem[{{Dalgarno} \& {McCray}(1972)}]{1972ARA&A..10..375D}
{Dalgarno}, A., \& {McCray}, R.~A. 1972, \araa, 10, 375

\bibitem[{{Dawson}(2013)}]{2013PASA...30...25D}
{Dawson}, J.~R. 2013, \pasa, 30, 25

\bibitem[{{Devine} \& {Bally}(1999)}]{1999ApJ...510..197D}
{Devine}, D., \& {Bally}, J. 1999, \apj, 510, 197

\bibitem[{{Field}(1965)}]{Field}
{Field}, G.~B. 1965, ApJ, 142, 531

\bibitem[{{Freyer} {et~al.}(2003){Freyer}, {Hensler}, \&
  {Yorke}}]{2003ApJ...594..888F}
{Freyer}, T., {Hensler}, G., \& {Yorke}, H.~W. 2003, \apj, 594, 888

\bibitem[{{Fujita} {et~al.}(2009){Fujita}, {Martin}, {Mac Low}, {New}, \&
  {Weaver}}]{2009ApJ...698..693F}
{Fujita}, A., {Martin}, C.~L., {Mac Low}, M.-M., {New}, K.~C.~B., \& {Weaver},
  R. 2009, ApJ, 698, 693

\bibitem[{{Heckman} {et~al.}(1990){Heckman}, {Armus}, \&
  {Miley}}]{1990ApJS...74..833H}
{Heckman}, T.~M., {Armus}, L., \& {Miley}, G.~K. 1990, \apjs, 74, 833

\bibitem[{{Hill} {et~al.}(2012){Hill}, {Joung}, {Mac Low}, {Benjamin},
  {Haffner}, {Klingenberg}, \& {Waagan}}]{Hill}
{Hill}, A.~S., {Joung}, M.~R., {Mac Low}, M.-M., {et~al.} 2012, \apj, 750, 104

\bibitem[{{Inoue} {et~al.}(2006){Inoue}, {Inutsuka}, \& {Koyama}}]{KoyamaII}
{Inoue}, T., {Inutsuka}, S.-i., \& {Koyama}, H. 2006, ApJ, 652, 1331

\bibitem[{{Jones} {et~al.}(2012){Jones}, {Stark}, \&
  {Ellis}}]{2012ApJ...751...51J}
{Jones}, T., {Stark}, D.~P., \& {Ellis}, R.~S. 2012, \apj, 751, 51

\bibitem[{{Joung} \& {Mac Low}(2006)}]{2006ApJ...653.1266J}
{Joung}, M.~K.~R., \& {Mac Low}, M.-M. 2006, \apj, 653, 1266

\bibitem[{{Kim} \& {Ostriker}(2015)}]{2015ApJ...802...99K}
{Kim}, C.-G., \& {Ostriker}, E.~C. 2015, \apj, 802, 99

\bibitem[{{Koyama} \& {Inutsuka}(2002)}]{Koyama}
{Koyama}, H., \& {Inutsuka}, S. 2002, ApJL, 564, L97

\bibitem[{{Leitherer} {et~al.}(1999){Leitherer}, {Schaerer}, {Goldader},
  {Delgado}, {Robert}, {Kune}, {de Mello}, {Devost}, \&
  {Heckman}}]{Leitherer1999}
{Leitherer}, C., {Schaerer}, D., {Goldader}, J.~D., {et~al.} 1999, \apjs, 123,
  3

\bibitem[{{Mac Low} \& {McCray}(1988)}]{1988ApJ...324..776M}
{Mac Low}, M.-M., \& {McCray}, R. 1988, \apj, 324, 776

\bibitem[{{Mac Low} {et~al.}(1989){Mac Low}, {McCray}, \&
  {Norman}}]{1989ApJ...337..141M}
{Mac Low}, M.-M., {McCray}, R., \& {Norman}, M.~L. 1989, \apj, 337, 141

\bibitem[{{Mandal} \& {Deshpande}(1994)}]{Mandal1994447}
{Mandal}, J., \& {Deshpande}, S. 1994, Computers \& Fluids, 23, 447

\bibitem[{{Mathis} {et~al.}(2002){Mathis}, {Whitney}, \& {Wood}}]{Mathis}
{Mathis}, J.~S., {Whitney}, B.~A., \& {Wood}, K. 2002, \apj, 574, 812

\bibitem[{{McCray} \& {Kafatos}(1987)}]{1987ApJ...317..190M}
{McCray}, R., \& {Kafatos}, M. 1987, \apj, 317, 190

\bibitem[{{McMillan}(2011)}]{2011MNRAS.414.2446M}
{McMillan}, P.~J. 2011, \mnras, 414, 2446

\bibitem[{{Mel{\'e}ndez} {et~al.}(2015){Mel{\'e}ndez}, {Veilleux}, {Martin},
  {Engelbracht}, {Bland-Hawthorn}, {Cecil}, {Heitsch}, {McCormick},
  {M{\"u}ller}, {Rupke}, \& {Teng}}]{2015ApJ...804...46M}
{Mel{\'e}ndez}, M., {Veilleux}, S., {Martin}, C., {et~al.} 2015, \apj, 804, 46

\bibitem[{{Melioli} {et~al.}(2013){Melioli}, {de Gouveia Dal Pino}, \&
  {Geraissate}}]{2013MNRAS.430.3235M}
{Melioli}, C., {de Gouveia Dal Pino}, E.~M., \& {Geraissate}, F.~G. 2013,
  \mnras, 430, 3235

\bibitem[{{Miyamoto} \& {Nagai}(1975)}]{1975PASJ...27..533M}
{Miyamoto}, M., \& {Nagai}, R. 1975, \pasj, 27, 533

\bibitem[{{Raymond} {et~al.}(1976){Raymond}, {Cox}, \&
  {Smith}}]{1976ApJ...204..290R}
{Raymond}, J.~C., {Cox}, D.~P., \& {Smith}, B.~W. 1976, \apj, 204, 290

\bibitem[{{Roy} {et~al.}(2013){Roy}, {Nath}, {Sharma}, \&
  {Shchekinov}}]{2013MNRAS.434.3572R}
{Roy}, A., {Nath}, B.~B., {Sharma}, P., \& {Shchekinov}, Y. 2013, \mnras, 434,
  3572

\bibitem[{{Rupke} {et~al.}(2002){Rupke}, {Veilleux}, \&
  {Sanders}}]{2002ApJ...570..588R}
{Rupke}, D.~S., {Veilleux}, S., \& {Sanders}, D.~B. 2002, \apj, 570, 588

\bibitem[{{Rupke} {et~al.}(2005){Rupke}, {Veilleux}, \&
  {Sanders}}]{2005ApJS..160..115R}
---. 2005, \apjs, 160, 115

\bibitem[{{Rybicki} \& {Lightman}(1986)}]{1986rpa..book.....R}
{Rybicki}, G.~B., \& {Lightman}, A.~P. 1986, {Radiative Processes in
  Astrophysics}

\bibitem[{{Sarazin}(1986)}]{1986RvMP...58....1S}
{Sarazin}, C.~L. 1986, Reviews of Modern Physics, 58, 1

\bibitem[{{Scarlata} \& {Panagia}(2015)}]{2015ApJ...801...43S}
{Scarlata}, C., \& {Panagia}, N. 2015, \apj, 801, 43

\bibitem[{{Shapiro} {et~al.}(1994){Shapiro}, {Giroux}, \&
  {Babul}}]{1994ApJ...427...25S}
{Shapiro}, P.~R., {Giroux}, M.~L., \& {Babul}, A. 1994, \apj, 427, 25

\bibitem[{{Sharma} {et~al.}(2014){Sharma}, {Nath}, {Chattopadhyay}, \&
  {Shchekinov}}]{2014MNRAS.441..431S}
{Sharma}, M., {Nath}, B.~B., {Chattopadhyay}, I., \& {Shchekinov}, Y. 2014,
  \mnras, 441, 431

\bibitem[{{Shopbell} \& {Bland-Hawthorn}(1998)}]{1998ApJ...493..129S}
{Shopbell}, P.~L., \& {Bland-Hawthorn}, J. 1998, \apj, 493, 129

\bibitem[{{Silich} {et~al.}(1996){Silich}, {Franco}, {Palous}, \&
  {Tenorio-Tagle}}]{1996ApJ...468..722S}
{Silich}, S.~A., {Franco}, J., {Palous}, J., \& {Tenorio-Tagle}, G. 1996, \apj,
  468, 722

\bibitem[{{Stone} {et~al.}(2008){Stone}, {Gardiner}, {Teuben}, {Hawley}, \&
  {Simon}}]{Stone-Athena}
{Stone}, J.~M., {Gardiner}, T.~A., {Teuben}, P., {Hawley}, J.~F., \& {Simon},
  J.~B. 2008, \apjs, 178, 137

\bibitem[{{Strickland} \& {Heckman}(2009)}]{2009ApJ...697.2030S}
{Strickland}, D.~K., \& {Heckman}, T.~M. 2009, \apj, 697, 2030

\bibitem[{{Strickland} {et~al.}(2002){Strickland}, {Heckman}, {Weaver},
  {Hoopes}, \& {Dahlem}}]{2002ApJ...568..689S}
{Strickland}, D.~K., {Heckman}, T.~M., {Weaver}, K.~A., {Hoopes}, C.~G., \&
  {Dahlem}, M. 2002, \apj, 568, 689

\bibitem[{{Strickland} {et~al.}(1997){Strickland}, {Ponman}, \&
  {Stevens}}]{1997A&A...320..378S}
{Strickland}, D.~K., {Ponman}, T.~J., \& {Stevens}, I.~R. 1997, \aap, 320, 378

\bibitem[{{Strickland} \& {Stevens}(2000)}]{StricklandStevens}
{Strickland}, D.~K., \& {Stevens}, I.~R. 2000, MNRAS, 314, 511

\bibitem[{{Suchkov} {et~al.}(1994){Suchkov}, {Balsara}, {Heckman}, \&
  {Leitherer}}]{1994ApJ...430..511S}
{Suchkov}, A.~A., {Balsara}, D.~S., {Heckman}, T.~M., \& {Leitherer}, C. 1994,
  \apj, 430, 511

\bibitem[{{Suchkov} {et~al.}(1996){Suchkov}, {Berman}, {Heckman}, \&
  {Balsara}}]{1996ApJ...463..528S}
{Suchkov}, A.~A., {Berman}, V.~G., {Heckman}, T.~M., \& {Balsara}, D.~S. 1996,
  \apj, 463, 528

\bibitem[{{Sutherland} \& {Bicknell}(2007)}]{SutherlandBicknell}
{Sutherland}, R.~S., \& {Bicknell}, G.~V. 2007, ApJS, 173, 37

\bibitem[{{Sutherland} \& {Dopita}(1993)}]{1993ApJS...88..253S}
{Sutherland}, R.~S., \& {Dopita}, M.~A. 1993, \apjs, 88, 253

\bibitem[{{Tenorio-Tagle}(1979)}]{TenorioTagle79}
{Tenorio-Tagle}, G. 1979, \aap, 71, 59

\bibitem[{{Tenorio-Tagle} {et~al.}(1999){Tenorio-Tagle}, {Silich}, {Kunth},
  {Terlevich}, \& {Terlevich}}]{1999MNRAS.309..332T}
{Tenorio-Tagle}, G., {Silich}, S.~A., {Kunth}, D., {Terlevich}, E., \&
  {Terlevich}, R. 1999, \mnras, 309, 332

\bibitem[{{Veilleux} {et~al.}(2005){Veilleux}, {Cecil}, \&
  {Bland-Hawthorn}}]{2005ARA&A..43..769V}
{Veilleux}, S., {Cecil}, G., \& {Bland-Hawthorn}, J. 2005, \araa, 43, 769

\bibitem[{{Veilleux} {et~al.}(1994){Veilleux}, {Cecil}, {Bland-Hawthorn},
  {Tully}, {Filippenko}, \& {Sargent}}]{1994ApJ...433...48V}
{Veilleux}, S., {Cecil}, G., {Bland-Hawthorn}, J., {et~al.} 1994, \apj, 433, 48

\bibitem[{{Vieser} \& {Hensler}(2000)}]{2000Ap&SS.272..189V}
{Vieser}, W., \& {Hensler}, G. 2000, \apss, 272, 189

\bibitem[{{Vieser} \& {Hensler}(2007)}]{2007A&A...472..141V}
---. 2007, \aap, 472, 141

\bibitem[{{Walter} {et~al.}(2002){Walter}, {Weiss}, \&
  {Scoville}}]{2002ApJ...580L..21W}
{Walter}, F., {Weiss}, A., \& {Scoville}, N. 2002, \apjl, 580, L21

\bibitem[{{Weaver} {et~al.}(1977){Weaver}, {McCray}, {Castor}, {Shapiro}, \&
  {Moore}}]{1977ApJ...218..377W}
{Weaver}, R., {McCray}, R., {Castor}, J., {Shapiro}, P., \& {Moore}, R. 1977,
  \apj, 218, 377

\bibitem[{{Williamson} {et~al.}(2014){Williamson}, {Thacker}, {Scannapieco}, \&
  {Br{\"u}ggen}}]{Williamson11062014}
{Williamson}, D.~J., {Thacker}, R.~J., {Scannapieco}, E., \& {Br{\"u}ggen}, M.
  2014, \mnras, 441, 389

\bibitem[{{Wofford} {et~al.}(2013){Wofford}, {Leitherer}, \&
  {Salzer}}]{2013ApJ...765..118W}
{Wofford}, A., {Leitherer}, C., \& {Salzer}, J. 2013, \apj, 765, 118

\bibitem[{{Wolfire} {et~al.}(1995){Wolfire}, {Hollenbach}, {McKee}, {Tielens},
  \& {Bakes}}]{1995ApJ...443..152W}
{Wolfire}, M.~G., {Hollenbach}, D., {McKee}, C.~F., {Tielens}, A.~G.~G.~M., \&
  {Bakes}, E.~L.~O. 1995, \apj, 443, 152

\bibitem[{{W{\"u}nsch} {et~al.}(2011){W{\"u}nsch}, {Silich}, {Palou{\v s}},
  {Tenorio-Tagle}, \& {Mu{\~n}oz-Tu{\~n}{\'o}n}}]{2011ApJ...740...75W}
{W{\"u}nsch}, R., {Silich}, S., {Palou{\v s}}, J., {Tenorio-Tagle}, G., \&
  {Mu{\~n}oz-Tu{\~n}{\'o}n}, C. 2011, \apj, 740, 75

\end{thebibliography}
\end{document}